\documentclass[aps,pra,twocolumn,showpacs,amsmath,amssymb,amsfonts,superscriptaddress,floatfix]{revtex4}
\usepackage{graphicx}
\graphicspath{{images/}}
\usepackage{bm}
\usepackage[dvips]{color}

\usepackage{subfig}
\usepackage{psfrag}

\usepackage{textcomp}

\newcommand{\comment}[1]{}

\newcommand{\tst}{\textstyle}

\newcommand{\mbf}{\mathbf}
\newcommand{\mrm}{\mathrm}
\newcommand{\ud}{\mathrm{d}}

\usepackage{natbib}         

\begin{document}

\author{Hauke Doerk-Bendig}
\affiliation{University of Ulm, Albert-Einstein-Allee 11, D-89069 Ulm, Germany}
\affiliation{Max-Planck-Institut f\"ur Plasmaphysik, Boltzmannstr. 2, D-85748 Garching, Germany}
\author{Zbigniew Idziaszek}
\affiliation{Institute of Theoretical Physics, Faculty of Physics, University of Warsaw, Ho\.{z}a 69, 00-681 Warsaw, Poland}
\author{Tommaso Calarco}
\affiliation{University of Ulm, Albert-Einstein-Allee 11, D-89069 Ulm, Germany}
\affiliation{ECT, I-38050 Villazzano (TN), Italy}

\title{Atom-ion quantum gate}

\begin{abstract}
We study ultracold collisions of ions with neutral atoms in traps.
Recently, ultracold atom-ion systems are becoming available in experimental setups, where their quantum states can be coherently controlled.
This allows for an implementation of quantum information processing combining the advantages of charged and neutral particles.
The state-dependent dynamics that is a necessary ingredient for quantum computation schemes is provided in this case by the short-range interaction forces depending on hyperfine states of both particles.

In this work we develop a theoretical description of spin-state-dependent trapped atom-ion collisions
in the framework of a Multichannel Quantum Defect Theory (MQDT) and formulate an effective single
channel model that reduces the complexity of the problem.

Based on this description we simulate a two-qubit phase gate between a $^{135}$Ba$^+$ ion and a
$^{87}$Rb atom using a realistic combination of the singlet and triplet
scattering lengths.
We optimize and accelerate the gate process
with the help of optimal control techniques. Our result is a gate fidelity $1-10^{-3}$ within
$350$\hbox{\textmu}s .

\end{abstract}

\pacs{34.50.cx,37.90.+j,03.67.Bg,03.67.Lx}

\maketitle
\section{introduction}
Ongoing developments in quantum information processing stimulate an intense
search for physical systems suitable for its implementation. Beside
solid-state and photonic systems, cold ions and neutral atoms represent
major candidates in this direction.

Neutral atoms can be accurately manipulated in dipole traps
\cite{Bergamini:04,Meschede:06}, optical lattices \cite{Bloch:02} or on atom chips
\cite{microtraps1,microtraps2} \comment{(TODO: more recent references on these
fields)}. Advanced evaporative and laser-cooling techniques allow their
preparation in the vibrational ground state of different trapping
potentials.
Single ions can be confined in Paul or Penning traps \cite{iontraps} and
sideband laser cooling allows to cool them down to the trap ground state.

Ultracold systems combining ions and neutral atom are currently being
explored \cite{AtomIonCloud,Vuletic,ZbyszekMQDT}.
Besides several new quantum mechanical aspects of this system, the studies
are motivated by potential applications. For example techniques of
sympathetic cooling of trapped atoms by laser-cooled trapped ions can
be developed
\cite{sympatheticCoolingOfAtomsByIons,sympatheticCoolingOfIonsByAtoms}.
In this paper we propose a scheme for a quantum gate that combines the
advantages of atoms and ions for quantum computation.

The trapping potentials of atoms and ions, although both built up with oscillating electromagnetic
fields, do not interfere with each other, since the oscillation frequencies of the respective fields typically
differ by orders of magnitude. The strength of the effective ion potential can be much stronger
than for neutral atoms. Tight confinement enables fast transport and together with
the good addressability of single trapped ions with lasers, this is among the advantages
of ions for implementing quantum computation.

Realization of the Mott insulator phase allows to prepare an array of atoms with well controlled number of particles in a single site of an optical lattice. In comparison to ions

The possibility to prepare an array of atoms in the Mott insulator phase in an optical lattice combined with the long decoherence times of neutral atoms is a reason to use atoms for the storage of quantum information. Furthermore, the two-particle interaction of an atom and an ion is typically much stronger than for two neutral atoms, which allows fast gate operations.

While qubits can be stored in internal electronic degrees of freedom
of both kinds of particles, the state dependent
dynamics suitable for two-qubit gates requires engineering of the two particle interaction. To this end one can use external
electromagnetic fields, e.g. magnetic Feshbach resonances that allow for a precise tuning of the two-body effective
scattering properties. The long-range atom-ion interaction also supports a trap-induced type of resonance \cite{ZIAtomIon},
because of their
generally state dependent nature, they will constitute a basic element of our quantum computation scheme.

In this work we solely make use of these trap-induced shape resonances that occur at relatively large distances. In this way we avoid some possible unwanted processes that may result from molecular dynamics at short distances. We nevertheless plan to include magnetic Feshbach resonances to our theory, which can be applied to perform two-qubit operations in a controlled collisions \cite{CalarcoFeshbach,Krych}.

\begin{figure}[ht]
\psfrag{atoms}{atoms in optical lattice} \psfrag{ion}{ion in rf trap}
\includegraphics[width=7cm]{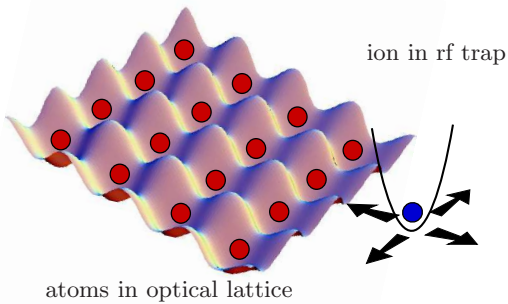}
\caption[Concept for Quantum Information with Atoms and Ions]{(Color online) Concept for quantum computation with
atoms and ions: Atoms are prepared in an optical lattice in a Mott insulator phase. A movable ion
entangles the atoms and can also be used for sympathetic cooling.\label{Fig:OLAtomAndIon}}
\end{figure}

A possible setup for quantum computation is schematically depicted in
Figure~\ref{Fig:OLAtomAndIon}. Atoms are stored in an optical lattice in a Mott insulator phase
such that each lattice site is occupied by exactly one atom. One movable ion is used to create
long-distance entanglement between pairs of atoms and to perform quantum gates. The basic ingredient of this
idea is the controlled and qubit-sensitive interaction between atoms
and ions. In this paper we focus on the dynamics of a single atom interacting with a single ion, nevertheless; our approach can be easily extended to the situation of many atoms, or more than one ion, at a later stage.

In this work we develop a theoretical model for spin-dependent atom-ion collisions.
For the case of an alkaline-earth ion and an alkali atom we formulate a model based on the Multichannel Quantum Defect Theory (MQDT) \cite{ZbyszekMQDT} taking into account the presence of trapping potentials \cite{GaoTrap}. Within our model the atom-ion interaction is described
by the long-range $1/r^4$ polarization potential combined with a set of quantum-defect parameters representing the effect of the short-range potential. The essential parameters for our approach are the singlet and triplet scattering lengths, which are not yet known with sufficient accuracy, but can probably be measured in upcoming experiments.
In our paper we discuss several different regimes for  the values of the singlet and the triplet scattering length. For some specific range of scattering lengths we are able to reduce the complexity of the problem by employing an effective single-channel model for the atom-ion dynamics.

Using the effective single-channel description we are able to simulate a two-qubit phase gate for an arbitrary
combination of atom and ion species. Applicability of the model, however, requires values of singlet and triplet scattering lengths that are nearly equal. In this case, within the single-channel description and for a specific system of $^{135}$Ba$^+$ ion and
$^{87}$Rb atom, we develop a phase gate process yielding a fidelity of $1-10^{-3}$ within the gate time
of $346$ \hbox{\textmu}s. Being equivalent to the CNOT gate, the phase gate is  universal for quantum computation \cite{Barenco}.
Therefore we demonstrate the feasibility of quantum computation
on the system under consideration. In a general situation, when the scattering lengths are not similar, the phase gate can be even faster; however this requires going beyond the single-channel effective description and is outside the scope of the present paper.

There are two main mechanisms that could lead to a failure of the quantum gate. One is the radiative
charge transfer, which in our scheme leads to a loss of both particles in case of heteronuclear species. In contrast, for a homonuclear collision \cite{Vuletic} the charge transfer results in a physically equivalent situation and therefore cannot be considered as a loss mechanism. Heteronuclear alkaline-earth ion -- alkali atom systems have the advantage of a relatively simple electronic level structure, and for the systems studied so far; Na-Ca$^+$ \cite{sympatheticCoolingOfIonsByAtoms,ZbyszekMQDT} and Rb-Ba$^+$ \cite{BaRbI}, the charge exchange rate remains much smaller than the elastic collisional rate, even in the presence of resonances. The second loss process results from spin changing collisions. In our scheme the qubits are encoded in hyperfine spin states, and collisions leading to final states outside of the computational basis have to be avoided. In the regime
of applicability of our single-channel effective model, the coupling between different channels is by definition very weak and those kinds of losses can be safely neglected. Even in a general situation, anyway, a multichannel treatment including all possible spin-state channels offers the possibility to gain control over spin-changing processes by appropriate engineering of the gate dynamics.

The paper is organized as follows. In Sec~\ref{Sec:setup} we describe the basic setup and model
used throughout the paper. Further we briefly discuss the atom-ion polarization interaction and
we introduce the concepts of correlation diagrams and trap induced resonances. The MQDT for
trapped particles as well as its reduction to a single-channel model is developed in
Sec~\ref{Sec:multichannelQDT}. The presented theory allows for computation  of eigenstates and eigenenergies either in single-channel or multi-channel situations. The dynamics of atom-ion collisions is discussed in
Sec.~\ref{Sec:AIdynamics}, where correlation diagrams and the Landau Zener theory help to
understand the features of the system. Sec.\ref{Sec:quantum_gate} presents the concepts and results
of our two-qubit phase gate simulations. A summary of our results, together with further perspectives and ideas,
is given in Sec~\ref{Sec:Outlook}.

\section{Basic Setup and Model}
\label{Sec:setup}
We consider a system consisting of a single atom and a single ion, stored in their respective
trapping potentials. Such potentials can be created with rapidly oscillating (rf) electric fields
for ions and with optical traps based on the ac Stark effect for atoms. These traps can be well
approximated as effective time-independent harmonic traps, as long as the particles are close to
the ground state of the potential. For this setup we introduce an effective Hamiltonian
\begin{equation}
\label{Eq:hamiltonian_general}
\begin{split}
H=&- \frac{\hbar^2}{2m_i} \Delta_i - \frac{\hbar^2}{2m_a} \Delta_a + \frac 1 2 m_i \omega_i^2(\mathbf{r}_i-\mathbf{d}_i)^2 \\
& + \frac 1 2 m_a \omega_a^2(\mathbf{r}_a-\mathbf{d}_a)^2 + W(|\mathbf{r}_i-\mathbf{r}_a|),
\end{split}
\end{equation}
where $m_{i(a)}$ is the mass of the ion (atom), $\omega_{i(a)}$ and $\mbf{d}_{i(a)}$ denote
frequency and location of the harmonic trapping potential of the ion (atom) and $W(r)$ is the
interaction potential. A microscopic derivation of the Hamiltonian Eq.~\eqref{Eq:hamiltonian_general}
can be found in \cite{ZIAtomIon}. Here, for simplicity we have assumed spherically symmetric
trapping potentials and the same trapping frequencies for atom and ion: $\omega_i=\omega_a=\omega$.
We stress, however, that our approach can be easily generalized to anisotropic trapping potentials
and different trapping frequencies \cite{ZIAtomIon}.
A general treatment would imply coupled center-of-mass (CM) and relative degrees of freedom, thus a six dimensional equation, but there are no fundamental difficulties.
In experiments both traps can be designed to
be spherically symmetric, while the assumption of the same trapping frequencies allows us to
decouple the relative and CM motions, thereby reducing the dimensionality of the
problem from six to three.
This choice simplifies our numerical calculations, on the other hand it
allows to capture the most important features of the system.

We transform the Hamiltonian Eq.~\eqref{Eq:hamiltonian_general}, introducing CM and relative coordinates, $\mathbf{R}_{\rm CM}=(m_i \mathbf{r}_i+m_a \mathbf{r}_a)/(m_i+m_a)$ and $\mathbf{r}=\mathbf{r}_i-\mathbf{r}_a$, respectively. Without loosing generality we can choose the coordinate frame such that the vector of trap separation $\mathbf{d}=\mathbf{d}_i-\mathbf{d}_a=d\,\mathbf{e}_z$
points in the $z$ direction. In this way we obtain the relative Hamiltonian
\begin{align}
\label{Eq:relativeHamiltonian2} {H}_{\rm rel}^{(d)}={H}_{\rm
rel}^{(0)}+\frac{1}{2}\mu\omega^2 d^2-\mu \omega^2 d z,\\
\intertext{where $\mu=m_im_a/(m_i+m_a)$ denotes the reduced mass of the atom-ion system and}
\label{Eq:relativeHamiltonian}
{H}_{\mathrm{rel}}^{(0)}=-\frac{\hbar^2}{2\mu}\mrm{\Delta}_r+\frac{1}{2}\mu\omega^2 r^2 +W(r)
\end{align}
is the Hamiltonian for the special case $\mrm{d}=0$.
\subsection{Atom-ion interaction}

\begin{figure}[ht]
\psfrag{atom}{Polarized atom} \psfrag{Ion}{Ion} \psfrag{rmin}{$r_{\rm min}$} \psfrag{Rast}{$R^{\ast}$} \psfrag{V(r)}{$V(r)\equiv-C_4/r^4$} \psfrag{r}{$r$} \psfrag{W(r)}{$W(r)$}
\includegraphics[width=7cm]{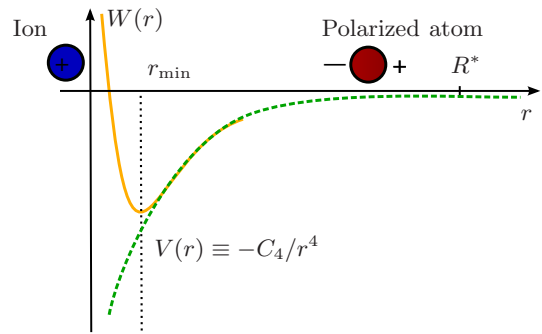}
\caption[Atom-ion interaction potential]{(Color online) The long-range part atom-ion interaction potential equals $-C_4/r^4$. At distances smaller than the potential minimum $r_{\rm min}$, repulsive terms start to dominate. Quantum defect theory replaces the actual potential $W(r)$ (solid line) with a reference potential $V(r)$ (dashed line) and includes the short-range effects using a quantum defect parameter related to the short range phase of the relative wavefunction. The characteristic range $R^*$ of the interaction is typically much larger than $r_{\rm min}$.} \label{Fig:potential}
\end{figure}

At large distances the atom-ion interaction potential has the asymptotic behavior $W(r)\simeq
-C_4/r^4$ ($r \rightarrow \infty$). This results from the fact that the ion charge polarizes the
electron cloud of the atom, and the induced dipole and the ion attract each other. Therefore, the
atom-ion interaction falls into an intermediate category, between the long-range Coulomb forces
$W(r) \sim 1/r$ and the van der Waals forces $W(r)\sim 1/r^6$ for neutral atoms. The interaction
constant $C_4$ can be expressed in terms of the electric dipole polarizability $\alpha$ of the atom
in the electronic ground state (S-state): $C_4 =\alpha e^2/2$. The electron charge is denoted as
$e$.
At short distances the interaction is dominated by the exchange forces, and higher order dispersion
terms ($C_6/r^6$, $C_8/r^8$) also become relevant. In our approach we model the short-range part of the potential using the quantum-defect
method, that is we do not require the knowledge of the exact form of the short-range interaction.
The interaction potential, in addition to the model potential, is depicted schematically in
Fig.~\ref{Fig:potential}.

By equating the interaction potential $C_4/{R^*}^4$ to the kinetic energy $\hbar^2/2\mu {R^*}^2$ we
can define some characteristic range
$R^*=\sqrt{2\mu C_4/\hbar^2}$
and corresponding characteristic energy $E^*=\hbar^2/2\mu{R^*}^2$ of the atom-ion interaction. Table
\ref{Tab:charUnits} gives the characteristic range and energy for some example atom-ion systems.
For comparison it also includes the harmonic oscillator length $l_0=\sqrt{\hbar/\mu\omega}$ for
$\omega_i = \omega_a = 2\pi\times 100$kHz.
\begin{table}
\begin{ruledtabular}
\begin{tabular}{r|ccc}
Atom-Ion System & $R^{\ast}(a_0)$ & $l_0(a_0)$ & $E^{\ast}/h$(kHz)\\
\hline $^{135}$Ba$^+$ + $^{87}$Rb & 5544 & 826 & 1.111\\    
$^{40}$Ca$^+$ + $^{87}$Rb & 3989 & 1178 & 4.142\\                
$^{40}$Ca$^+$ + $^{23}$Na & 2081 & 1572& 28.545\\                 
\end{tabular}
\end{ruledtabular}
\caption{\label{Tab:charUnits} Characteristic length and energy scale for example systems.
Oscillator lengths are calculated with $\omega_i = \omega_a = 2\pi\times 100$kHz.}
\end{table}

\subsection{Single-channel quantum defect treatment}

The short range interaction potential between atom and ion is typically quite complicated and in
most cases it is not known theoretically with an accuracy sufficient to determine the
scattering properties in the limit of ultracold energies. In order to avoid complications while
using the explicit form of the short-range potentials, we resort to the quantum-defect method,
which allows to include the effects of the short-range forces in an effective way. This consists in
substituting the actual potential by the reference potential $V(r)=-C_4/r^4$ at all distances (see Fig.~\ref{Fig:potential}), and
assigning an appropriate short-range phase to the wavefunction to model the effects of the
short-range potential.

We illustrate this approach by solving the relative Schr\"odinger equation ${H}_{\rm
rel}^{(0)}\Psi(\mbf{r})=E\Psi(\mbf{r})$ at $d=0$. To this end we apply the partial wave
decomposition
\begin{equation}
\Psi(\mbf{r})=\sum_{lm}Y_{lm}(\hat{\bf r})\psi_l(r)/r,
\end{equation}
to obtain the radial Schr\"odinger equation
\begin{equation}
\label{Eq:singleRadial}
\left[-\frac{\hbar^2}{2\mu} \frac{\partial^2}{\partial
r^2}+ \frac{\hbar^2}{2\mu}\frac{l(l+1)}{r^2}+\frac{\mu
\omega^2}{2} r^2 +V(r) -E \right]\psi_{l}(r) = 0
\end{equation}
for the radial wavefunctions $\psi_l(r)$. Here, $Y_{lm}$ are the spherical harmonic functions describing the angular part of the 3D wavefunction, where $l$ and $m$ are the quantum numbers 
of the relative angular momentum and its projection on the symmetry axis $z$, respectively.
In the limit of $r\rightarrow 0$ we can neglect trapping potential, energy and centrifugal barrier in comparison to $V(r)=-C_4/r^4$, which yields
\begin{equation}
\label{Eq:singleRadialLim}
\left[-\frac{\hbar^2}{2\mu} \frac{\partial^2}{\partial
r^2}-\frac{C_4}{r^4}\right]\psi_{l}(r)=0,
\end{equation}
with the solution
\begin{equation}
\label{Eq:psi_l0E0} \psi_l(r)=r \sin\left(\frac{R^{\ast}}{r}+\varphi\right),\quad r \rightarrow 0
\end{equation}
where $\varphi$ is a parameter that can be interpreted as the short range phase. In our method we
treat Eq.~\eqref{Eq:psi_l0E0} as a boundary condition that we impose on the radial wave functions
at short distances, while solving the relative Schr\"odinger equation in the general case $d \neq
0$.

In the absence of a trapping potential, and for $l=0$ and $E=0$, the solution Eq.~\eqref{Eq:psi_l0E0}
becomes valid at all distances. By comparing the long range behavior of Eq.~\eqref{Eq:psi_l0E0}:
$\lim_{r\rightarrow \infty}\psi_0(r)/r\sim(1+R^\ast/r\,\cot\varphi)$, with the well known asymptotic
form of the $s$-wave radial wave function: $\lim_{r\rightarrow \infty}\psi_0(r)/r \sim 1-a/r$, we can
relate the short range phase to the scattering length
\begin{equation}
\label{Eq:srphase_scatteringlenght} a=- R^\ast \cot\varphi,
\end{equation}
which is a measurable physical quantity. We note that $R^\ast$ determines the typical length-scale
for the scattering length. In this chapter we have focused only on single-channel
collisions, not considering internal states of the particles. The present approach is generalized in
Sec.~\ref{Sec:multichannelQDT} to the realistic multichannel situation.
\subsection{Correlation diagram and trap induced resonances}
\label{Sec:CorrDiag}
\begin{figure}[ht]
\includegraphics[width=7cm]{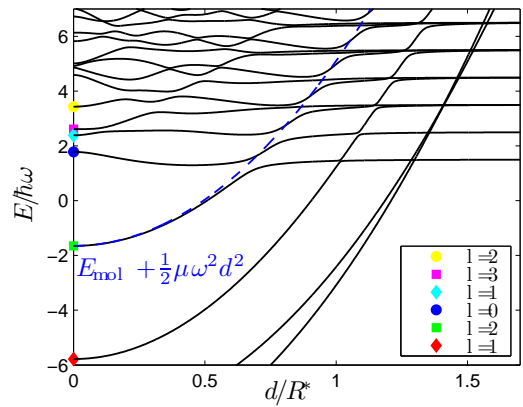}
\caption[Correlation Diagram]{(Color online) An example correlation diagram calculated for $R^*=3.68 l_0$ and the
short-range phase $\varphi=0.74\pi$ showing the energy spectrum versus the trap separation. The
partial wave number $l$ is given for the lowest states at $d=0$. As explained in the text, molecular state energies have an
approximate parabolic $d$-dependence indicated by the dashed line.} \label{Fig:exampleSpectrum}
\end{figure}
\begin{figure}[ht]
\centering \psfrag{E}{$E$} \psfrag{trap}{$\frac 1 2 \mu \omega^2 (\mathbf{r}-\mathbf{d})^2$}
\psfrag{potential}{$-\frac{C_4}{r^4}$} \psfrag{r}{$r$} \psfrag{d}{$d_{\rm res}$} \psfrag{molecular
state}{\small molecular state} \psfrag{vibrational state}{\small vibrational state}
\includegraphics[width=7cm]{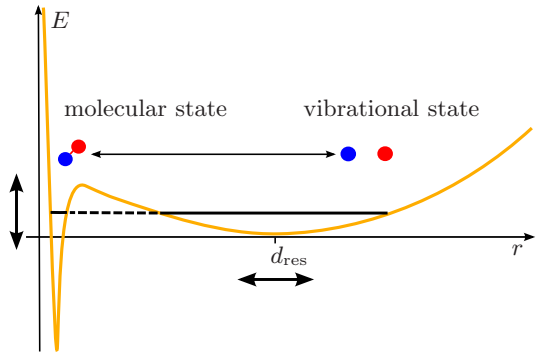}
\caption[Trap-Induced Resonance]{(Color online) Trap-induced shape resonance: at a certain trap separation $d=d_{\rm res}$, the energy of a molecular bound state becomes degenerate to a trap vibrational energy. The adiabatic eigenenergies exhibit an avoided crossing at this position. The arrows indicate that the molecular energy is shifted, if the relative trap position is changed.} \label{Fig:ShapeResonance}
\end{figure}

In order to obtain an intuitive understanding of atom-ion collisions, we describe them in terms of correlation diagrams, showing the energy spectrum as a function of the trap separation $d$ (see Fig. \ref{Fig:exampleSpectrum}). Such correlation
diagrams in our case connect the  asymptotic vibrational states for large trap separation to the molecular and vibrational
states at zero trap separation. At large distances we find harmonic oscillator-like equidistant
eigenenergies that are independent of $d$. Molecular bound states, that corresponds to the
eigenstates with energies well below the zero-point vibration energy $E_0 = 3/2 \hbar \omega$,
experience a quadratic shift with the distance $d$. This can be easily understood by noting that
the bound states $\Psi_\mrm{mol}(r)$ are well localized around $r = 0$, and $\langle
\Psi_\mrm{mol}| H_\mrm{rel}^{(d)} |\Psi_\mrm{mol}\rangle \approx E_\mrm{mol}+\frac{1}{2} \mu
\omega^2 d^2$, where $E_{\rm mol}$ is the molecular binding energy at $d=0$.
Beside the given arguments the quadratic shift in the molecular energy becomes immediately clear in Fig.~\ref{Fig:ShapeResonance}. The molecular potential `hangs' from the relative trapping potential in the low-distance region, thus increasing (decreasing) $d$ shifts up (down) the molecular energy as $d^2$.

At some particular distances, the energies of the molecular states become equal to the energies of
the vibrational levels (see Fig.~\ref{Fig:ShapeResonance}), and the spectrum exhibits avoided
crossings, known as the trap-induced shape resonances \cite{Stock}. By slowly changing the trap separation $d$ we can pass through the resonance adiabatically, converting the
trap vibrational states into molecular states, thus producing molecular ions. Since this process is
reversible, we can coherently control the dynamics of our system by appropriately adjusting the
trap distance.

\section{Quantum-defect theory for trapped particles}
\label{Sec:multichannelQDT}
\subsection{Multichannel formalism }
In general, the interaction properties depend on the internal state of two colliding particles. For these internal states we choose a convenient basis, in which the two-particle Hamiltonian is diagonal at large particle distance, where the interaction potential is negligible.
We then refer to the two-particle basis states as scattering channels. The wavefunction is decomposed into the chosen basis, which allows us to write down Schr\"odinger's equation in matrix form.
In the following we introduce an MQDT formalism, following closely the formulation by F. Mies
\cite{MiesMQDT}, and adopting it to a situation including an external trapping potential.
Assuming the
same trapping frequencies for atom and ion, the CM and relative degrees of freedom are decoupled.
In this case we can describe the relative motion with the close-coupled Schr\"odinger equation
\begin{equation}
\label{Eq:Schr} - \frac{\hbar^2}{2 \mu} \Delta \mbf{\Psi}(r) + \left[\mbf{W}(r)+ \mbf{U}(\mbf{r}) -E
\mbf{I}\right] \mbf{\Psi}(r) = 0.
\end{equation}
Here
$\mbf{I}$ denotes the identity
matrix, $\mbf{W}(r)$ is the interaction matrix, which is asymptotically diagonal
\begin{equation}
W_{ij}(r) \stackrel{r \rightarrow \infty}{\longrightarrow} \left[ E_i^{\infty} -  \frac{C_{4}}{r^4} \right]\delta_{ij},
\end{equation}
with $\{i,j\}$ indicating the channels. The trapping potential $\mbf{U}(r)$ is diagonal at all distances
\begin{equation}
U_{ij}(\mbf{r}) = \frac{1}{2} \mu \omega^2 (\mbf{r}-\mbf{d})^2 \delta_{ij}.
\end{equation}
The matrix $\mbf{\Psi}(r)$ contains $N$ linearly independent solutions, where $N$ is the number of
channels. The threshold energies for the molecular dissociation in channel $i$ are denoted by
$E_i^{\infty}$.

\subsubsection{Special case: $d=0$}
\label{Sec:multd0}

For $\mbf{d}=0$ the external potential is spherically symmetric and the dynamics for different
relative angular momenta $l$ is decoupled. We can decompose $\mbf{F}(r)$ into a partial wave
expansion
\begin{equation}
\mbf{\Psi}(\mbf{r}) = \sum_{l} c_l \mbf{F}_{l}(r) Y_{l0}(\mbf{\hat{r}})/r,
\end{equation}
where
Here, for simplicity we consider only the $m=0$ subspace.
The radial wave functions $\mbf{F}_{l}(r)$ fulfill
\begin{equation}
\label{Eq:RadialSchr} \left[H_l \,\mbf{I} + \mbf{W}(r) \right] \mbf{F}_l(r) = E \mbf{F}_l(r),
\end{equation}
with
\begin{equation}
H_l = - \frac{\hbar^2}{2 \mu} \frac{\partial^2}{\partial r^2} + \frac{\hbar^2 l(l+1)}{2 \mu r^2} + \frac12 \mu \omega^2 r^2.
\end{equation}
In our calculations we model the short-range potential by choosing appropriate short-range phases
$\varphi_i$ for each of the channels. This is equivalent to setting $W_{ij}(r) = \left(E_i^{\infty}
-  C_{4}/r^4 \right) \delta_{ij}$ at all distances. In this case the reference potentials that are
necessary to define the MQDT functions \cite{MiesMQDT} can be simply taken as diagonal elements of $\mbf{W}(r)$:
$V_i(l,r) \equiv W_{ii}(r) + \frac{\hbar^2 l(l+1)}{2 \mu r^2}$. Given the reference potentials,
$V_i(l,r)$ one can associate to them a pair of linearly independent solutions $\hat{f}_i(l,r)$ and
$\hat{g}_i(l,r)$ of the single-channel Schr\"odinger equation that have WKB-like normalization at
small distances
\begin{align}
\label{Eq:fghat} \left.
\begin{array}{lll}
\hat{f}_{i}(l,r) & \cong & k_i(l,r)^{-1/2} \sin \beta_i(l,r),\phantom{\Big(}\\
\hat{g}_{i}(l,r) & \cong & k_i(l,r)^{-1/2} \cos \beta_i(l,r),\phantom{\Big(}\\
\end{array}
\right\}\,\, r \sim r_\mrm{min},
\end{align}
where $k_i(l,r) = \sqrt{2 \mu (E - V_i(l,r)}/\hbar$ is the local wavevector and $\beta_i(l,r) =
\int^r\!\mrm{d}x\,k_i(l,x)$ is the WKB phase. Here, $r_\mrm{min}$ denotes a typical distance where
the minima of the realistic potential occur, and the semiclassical approximation is applicable. In
our modeling $r_\mrm{min} \rightarrow 0$, and Eq.~\eqref{Eq:fghat} describe the asymptotic behavior $r
\rightarrow 0$.

The solution to Eq.~\eqref{Eq:RadialSchr} can be expressed in terms of a pair of functions
$\hat{\mbf{f}}_l(r) \equiv \{\delta_{ij} \hat{f}_{i}(l,r)\}$ and $\hat{\mbf{g}}_l(r) \equiv
\{\delta_{ij} \hat{g}_{i}(l,r)\}$:
\begin{equation}
\label{Eq:Fl1} \mbf{F}_l(r) = \left[\hat{\mbf{f}}_l(r) + \hat{\mbf{g}}_l(r) \mbf{Y}_l(E)\right]
\hat{\mbf{A}}.
\end{equation}
Here, $\mbf{Y}_l(E)$ is the quantum-defect matrix that represents the effects of the short-range
potential, in particular couplings between channels, and will be discussed later.
The Matrix $\hat{\mbf{A}}$ has constant coefficients and is determined by the boundary conditions
at $r \rightarrow \infty$. We note that in MQDT the functions $\hat{\mbf{f}}_l(r)$ and
$\hat{\mbf{g}}_l(r)$ describe in general only the asymptotic $(r \rightarrow \infty)$ behavior of
$\mbf{F}_l(r)$. Due to our choice of $\mbf{W}(\mbf{r})$ and $V_i(l,r)$, however, in our case these functions
will be valid at all distances.

In analogy to MQDT in free space, we introduce another type of solutions, normalized at $r
\rightarrow \infty$. At large distances the harmonic potential dominates and the solution vanishing
at $r \rightarrow \infty$ reads
\begin{align}
\label{Eq:phi} \phi_i(l,r) \stackrel{r \rightarrow \infty}{\longrightarrow} D_\nu (\sqrt{2} r/\xi),
\end{align}
where $D_\nu(z)$ is the parabolic cylinder function, $E=E^\infty_i + \hbar \omega (\nu + \frac12)$,
and $\xi$ is the harmonic oscillator length $\xi=\sqrt{\hbar/\mu \omega}$. The two types of
solutions, Eq.~\eqref{Eq:fghat} and Eq.~\eqref{Eq:phi}, can be related by the MQDT functions $\nu_i(l,E)$
and ${\cal N}_i(l,E)$:
\begin{align}
\begin{split}
\label{Eq:f}
\phi_i(l,r) = {\cal N}_i(l,E) \Big[& \cos \nu_i(l,E) \hat{f}_i(l,r) \\
& - \sin \nu_i(l,E) \hat{g}_i(l,r) \Big].
\end{split}
\end{align}
The function $\nu_i(l,E)$ mixes the two solutions Eq.~\eqref{Eq:fghat} leading to the exponentially
decaying function $\phi_i(l,r)$, whereas ${\cal N}_i(l,E)$ provides the overall normalization. In
fact the normalization can be calculated directly from $\nu_i(l,E)$ \cite{MiesMQDT}:
\begin{equation}
{\cal N}_i(l,E) = \left(\frac{\hbar^2}{2 \mu} \frac{\partial \nu_i(l,E)}{\partial E}\right)^{-1/2}.
\end{equation}
Now, the wave function $\mbf{F}_l(r)$ can be equivalently expressed in terms of solutions
${\bm \Phi}_l(r) \equiv \{\delta_{ij} \phi_{i}^l(r)\}$ normalized at infinity: ,
\begin{equation}
\label{Eq:Fl2} \mbf{F}_l(r) = {\bm \Phi}_l(r)  \mbf{A}.
\end{equation}
By comparing Eq.~\eqref{Eq:Fl1} with Eq.~\eqref{Eq:Fl2} one arrives at the following equation:
\begin{equation}
\label{Eq:A} \left[\mbf{Y}_l(E)  + \tan {\bm \nu}_l(E) \right] {\bm N}_l(E) \cos {\bm \nu}_l(E)
\mbf{A} = 0,
\end{equation}
where ${\bm \nu}_l(E) \equiv \{\delta_{ij} \nu_{i}^l(E)\}$ and ${\bm N}_l(E) \equiv \{\delta_{ij}
{\cal N }_{i}^l(E)\}$. This has a nontrivial solution ($\mbf{A} \neq 0$), if
\begin{equation}
\label{Eq:Det} \left|\mbf{Y}_l(E) + \tan {\bm \nu}_l(E)\right| = 0,
\end{equation}
which is a standard condition determining bound states in the MQDT approach. From Eq.~\eqref{Eq:Det} one can
evaluate eigenenergies in the multichannel case, while the eigenstates are given by Eq.~\eqref{Eq:Fl2},
with $\mbf{A}$ determined from Eq.~\eqref{Eq:A}. 
This procedure yields a set of eigenfunctions $\mbf{F}_{ln}(r) = {\bm \Phi}_l(r) \mbf{b}_{ln}$ and
corresponding eigenenergies $E_{ln}$, where $\mbf{b}_{ln}$ is a constant vector, and the label $n$
enumerates the solutions:
\begin{equation}
\label{Eq:RadialSchr1} \Big(H_l \,\mbf{I} + \mbf{W}(r) \Big) \mbf{F}_{ln}(r) = E_{ln}
\mbf{F}_{ln}(r).
\end{equation}
Similarly to ordinary scalar wave functions, the multichannel eigenstates corresponding to
different non-degenerate eigenenergies are orthonormal:
\begin{equation}
\int_0^\infty \!\! \ud r \,\mbf{F}_{ln}(r)^\dagger \mbf{F}_{lm}(r) = \delta_{nm}.
\end{equation}

\subsubsection{Generalization to $d\neq0$}
\label{Sec:multdneq0}

At nonzero trap separation, the Hamiltonian is no longer rotationally invariant, and the procedure
presented in the previous section based on decoupling of states with different values of $l$ does not apply.
Nevertheless, we can utilize the previous solutions at $\mbf{d} =0$ to diagonalize the full problem Eq.~\eqref{Eq:Schr}
at  $\mbf{d} \neq 0$. To this end the total wave function is decomposed in terms of
arbitrary expansion coefficients $c_{ln}$:
\begin{equation}
\label{Eq:Fmulti} \mbf{F}(\mbf{r}) = \sum_{ln} c_{ln} \mbf{F}_{ln}(r) Y_{l0}(\mbf{\hat{r}})/r.
\end{equation}
Substituting this into Eq.~\eqref{Eq:Schr} and setting $\mbf{d}=d\mbf{e}_z$ we arrive at the set of
coupled equations
\begin{equation}
\left(E_{ln}+{\tst \frac12} \mu \omega^2 d^2 \right) c_{ln}
+ \mu \omega^2 d \sum_{l^\prime n^\prime} D^{l^\prime n^\prime}_{l n} c_{l^\prime n^\prime} = E c_{ln},
\end{equation}
which in principle can be solved numerically with standard methods for matrix diagonalization. Here
\begin{equation}
\label{Eq:MatrixelementMulti}
D^{l^\prime n^\prime}_{n l} \equiv \langle Y_{l0} |\cos \theta|
Y_{l^\prime0} \rangle \int_0^\infty \!\! \ud r\, \mbf{F}_{ln}(r)^\dagger r \mbf{F}_{ln^\prime}(r),
\end{equation}
and
\begin{equation}
\begin{split}
\langle Y_{l0} |\cos \theta| Y_{l^\prime0} \rangle = &
\int \!\! \ud \Omega\, Y_{l0}^\ast(\mbf{\hat{r}}) \cos \theta Y_{l^\prime0}(\mbf{\hat{r}}) \\
= & \frac{l+1}{\sqrt{(2l+1)(2l+3)}} \delta_{l+1,l^\prime} \\
& + \frac{l}{\sqrt{(2l-1)(2l+1)}} \delta_{l-1,l^\prime}.
\end{split}
\end{equation}

\subsubsection{Parametrization of $\mbf{Y}_l(E)$ and the frame transformation}
\label{Sec:frameTransformation}

In the regime of ultracold collisions, the variation of the total energy $E$ and the height of the angular momentum barrier (for the lowest partial waves which are important in the ultracold regime \cite{ZbyszekMQDT}) are much smaller than the depth of the potential at $r \sim r_\mrm{min}$ where the matrix $\mbf{Y}_l(E)$ is defined. Therefore it is justified to neglect the dependence of $\mbf{Y}_l(E)$ on
both energy and angular momentum, and to set $\mbf{Y}_l(E) \cong \mbf{Y}$. In this way, determining
the matrix $\mbf{Y}$ at a single value of energy, we may describe the atom-ion collisions in the
whole regime of ultracold temperatures.

In the paper we consider collisions of an alkali atom with an alkaline-earth ion in their electronic
ground states. Hence the asymptotic channel states that are used in the Schr\"odinger equation
Eq.~\eqref{Eq:RadialSchr}, can be characterized by the hyperfine quantum numbers $f_1$,$m_{f_1}$ and
$f_2$,$m_{f_2}$ for ion and atom respectively, and by the angular-momentum quantum numbers $l$ and
$m_l$ of the relative motion of the atom and ion CM. In the rest of this section we
label those channels by $\alpha = \{f_1 f_2 m_{f_1} m_{f_2} l m_l\}$.

At short distances, the potential matrix becomes diagonal in the molecular basis characterized by the total electron and nuclear spins and their projections, because the short-range forces depend on the electronic configuration of the entire atom-ion molecular complex. In fact, the molecular potentials that correlate with atom and ion electronic ground states at large distances depend only on the total electron spin $S$ \cite{ZbyszekMQDT}. For our choice of species the electronic configurations are identical as in collisions of two hydrogen atoms. Thus, $S$ can take the values $0$ (singlet configuration) and $1$ (triplet configuration). Hence the quantum defect matrix $\mbf{Y}$, which contains the full interaction information, can be parameterized with only two constants, the singlet $a_s$ and triplet $a_t$ scattering lengths. These constants depend only on the species.

In our approach we apply a frame transformation to find $\mbf{Y}$ in the basis of hyperfine states \cite{Gao,BurkePRL1998}. As shown in Ref.~\cite{ZbyszekMQDT}, this approximation is very accurate for atom-ion collisions due to a clear separation of length scales associated with the short-range and long-range forces. On the one hand, exchange interaction becomes significant only at distances of the order of few tens of $a_0$ (atomic units), when the electronic wavefunctions of atom and ion begin to overlap. On the other hand, the polarization forces are very long-ranged and they are modified by the presences of the centrifugal barrier only at large distances of the order of $R^\ast$.

\subsection{Reduction to an effective single-channel model in the case of $a_s \approx a_t$}
\label{Sec:SingleChannelModel}

The off-diagonal matrix elements in the quantum defect matrix $\mbf{Y}$ are proportional to the
"coupling" scattering length, that is defined as $1/a_c = 1/a_s - 1/a_t$ \cite{ZbyszekMQDT}.
Therefore in the case of $a_s \approx a_t$ the coupling between channels is weak and the many
channel description can be effectively reduced to a single-channel model. To this end we solve the
multi-channel problem at $d=0$, and we find the corresponding eigenenergies $E_{nl}$ and eigenstates
$\mbf{F}_{nl}(r)$ from Eq.~\eqref{Eq:A} and Eq.~\eqref{Eq:Det}. If the mixing between channels is weak, in each of the eigenfunction $\mbf{F}_{nl}(r)$, there is only one channel that dominates, i.e. the vector $\mbf{b}_{ln}$ has only one element which is close to unity. We divide the total multichannel spectrum into $N$ distinct subsets according to the channel that gives the leading contribution to $\mbf{F}_{nl}(r)$, and for each of the subset we determine the effective short-range phase $\varphi^\mrm{eff}_l$ (or the scattering
length $a_l^\mrm{eff} = - R^\ast \cot \varphi^\mrm{eff}_l$). This is done by matching the
multichannel spectrum $E_{nl}$ in each of the subsets with the single-channel generated by Eq.~\eqref{Eq:singleRadial} with the quantum-defect parameter $\varphi^\mrm{eff}_l$. In the limit of zero coupling between channels, the effective scattering length $a_l^\mrm{eff}$ is equal to $a_s=a_t$. In the presence of weak coupling, $a_l^\mrm{eff}$ is in general different from 
$a_s$ and $a_t$, since the asymptotic channels typically correlate both to singlet and triplet molecular states at small distances. This procedure yields a set of $N$ short
range phases $\{\varphi^\mrm{eff}_l\}$, which are used at a later stage in the single-channel
calculations. We note that the effective phases $\{\varphi^\mrm{eff}_l\}$ depend on the relative
angular momentum, and in principle they weakly depend on the energy. We have verified, however,
that within the considered range of energies limited to the bound states close to the dissociation
threshold, and to a few tens of the lowest vibrational states, the variations of
$\varphi^\mrm{eff}_l(E)$ with the energy are negligible.

For collisions when only a single open channel exists, the remaining closed channels are
typically only weakly coupled to the open channel (apart from the case of resonances), and the
resulting multichannel wave function is dominated by the open-channel component. The situation
changes, however, when there are more open channels, and the channel mixing can be significant. We
have investigated this issue numerically, picking the specific ion and atom pair $^{135}$Ba$^+$--$^{87}$Rb
and the trapping frequency $\omega_i=\omega_a=2\pi\times 30$kHz. We have considered
collisions within the $m_{F}=3$ subspace, assuming that initially the particles are prepared in the
channel $\alpha_1=\{f_i=1,m_{f_i}=1,f_a=2,m_{f_a}=2\}$. This choice is relevant for our modeling of
the quantum gate, as we show later. For the collision energies above the dissociation threshold of
the channel $\alpha_1$, a second open channel exists, with
$\alpha_2=\{f_i=2,m_{f_i}=2,f_a=1,m_{f_a}=1\}$. In this case we find that admixture of the two
remaining closed channels is negligible, whereas the contribution of both channels $\alpha_1$ and
$\alpha_2$ in the multichannel eigenstates is typically large, and the contributions from
$\alpha_1$ and $\alpha_2$ cannot be separated. The only exception is the case of similar $a_s$ and
$a_t$, where the inter-channel coupling $1/a_c$ is small, and the multichannel eigenstates are
dominated by the single-channel contributions.

We estimate the validity of the single channel approximation by calculating the overlap of the exact multichannel and the single channel
wave functions, in the range of trap separations $d$, interesting for our dynamics. The minimum over $d$ yields some overall fidelity related to the reduction to the single channel model. For similar singlet and triplet phases, we have  $\phi_l^\mrm{eff} \approx \phi_s \approx \phi_t$, and the multichannel wave functions differ from their single-channel counterparts by the presence of negligible contributions in the channels others than the dominating one. The relative contributions of individual channels are given by the vectors $\mbf{b}_{ln}$ that are obtained in the calculation of eigenstates at $d=0$. When $d>0$ one has to take into account that the multichannel wave functions are linear combinations of the solutions at $d=0$ (see Eq.~\eqref{Eq:Fmulti}), and the overlap between single- and multi-channel eigenstates is a linear combination of the overlaps calculated at $d=0$ with the expansion coefficients given by $|c_{ln}|^2$. 
%

Resonances occurring at $d=0$ can lead to significant channel mixing for few states, although the average fidelity is very high. The more these highly mixed states contribute to the wavefunctions, the larger is the error of the model. For the error estimation used in this work, we only use the trap ground state at maximal $d$ and a molecular state at minimal $d$ of interest. We find that the two fidelities lie in the same range, thus we take the minimum of them and assume the result as a lower bound for the fidelity at intermediate distances.
We have additionally verified that reductions to the single channel model works best for positive values of singlet and triplet scattering lengths
around $R^{\ast}$.

\section{Atom-ion dynamics}
\label{Sec:AIdynamics}
Traps for individual ultracold atomic particles used in schemes for quantum information processing provide in most cases for the ability to manipulate the particles' motion via appropriate tuning of external trap parameters. This is in particular the case for optical lattices and Paul traps, where field polarizations and intensities can be changed to control the shape and position of the traps to a high degree of accuracy. Our proposal relies on these standard techniques, thereby introducing the innovative aspect of combining traps for ions and atoms. As already discussed elsewhere \cite{ZIAtomIon}, the physical mechanisms generating the traps for ions and atoms are different and lead, under appropriate conditions, to independent microscopic traps, which can be modeled as follows.

In this chapter dynamics will be described by introducing a time dependent trap displacement $d(t)$. Below a certain distance $d\sim R^*$, the eigenenergies of the system start to depend on the spin state as well as on $d$ itself, and positions of trap-induced resonances are determined by the internal state of both particles. In this way trap displacement can be used for spin-dependent control of the atom-ion system.

\subsection{Landau-Zener Theory}
\label{Sec:LandauZener}
The Landau-Zener formula gives a basic understanding of the atom-ion dynamics in the vicinity of
trap-induced shape resonances. It describes a general two-level system whose eigenstates
$|\Psi_{1}\rangle$ and $|\Psi_2\rangle$ are coupled by some kind of interaction, and in which the
two eigenlevels $E_1$ and $E_2$ form an avoided crossing when varying an external parameter. In our case this external parameter is the trap displacement $d$. The probability of a nonadiabatic
passage of the crossing \cite{ZIAtomIon},
\begin{equation}
\label{Eq:LZprobability1} P_{\mathrm{na}}=\exp\left( -2\pi \frac{|\langle\Psi_1 |{H}|\Psi_2\rangle
|^2}{\hbar |\dot{d}\,\partial E_{12}/\partial d |}\right),
\end{equation}
depends on the coupling matrix element, the velocity $\dot{d}$ of passage of the resonance and the
relative slope $\partial E_{12}/\partial d$ of the levels, where $E_{12}=E_1-E_2$. A fast passage
of the avoided crossing ($P_{\mathrm{na}} \approx 1$) results in a nonadiabatic evolution, preserving the shape of the wave function. At small velocities the resonance is passed adiabatically ($P_{\mathrm{na}} \approx 0$), i.e. the system follows its eigenenergy curves.
In the trapped atom-ion system, for certain trap separations, the energy of some molecular bound states becomes equal to harmonic-oscillator energies, resulting in the avoided crossings. If such avoided crossing is passed adiabatically, then the initial harmonic-oscillator state with atom and ion located in their separated traps can evolve into a molecular bound state, where the atom and ion are trapped in a combination of the two external potentials. This process is reversible and we use it in Sec.~\ref{Sec:quantum_gate} to realize an entangling two-qubit operation.

In order to precisely predict the outcome of a collision process, in our simulations we have
calculated time evolution numerically, using the Landau-Zener formula only as a guide to estimate
the relevance of the avoided crossings for the transfer process. Since the energy of the molecular
bound states changes according to $E_\mathrm{{mol}}(d)\approx E_{\mathrm{mol}}(0)+1/2\mu\omega^2
d^2$ (see Fig.~\ref{Fig:exampleSpectrum}), deeply bound states can cross vibrational states only at
large $d$. In this case avoided crossings are very weak, due to the fact that $\Delta E$ decays
exponentially with the trap distance \cite{ZIAtomIon}. Hence, the deeply bound states have no
relevance for the dynamics and in our simulations we have included only shallow bound states that
are closest to the dissociation threshold.

\subsection{Full dynamics in the single-channel model}
In the case of similar $a_s$ and $a_t$, when the effective single-channel description is applicable, we
describe the dynamics of the controlled atom-ion collision with the following time-dependent
Hamiltonian:
\begin{equation}
H_{\mathrm{rel}}^{(d)}(t)=H_{\mathrm{rel}}^{(0)}+ \frac 1 2 \mu \omega^2 d(t)^2- \mu \omega^2 d(t)z.
\end{equation}
We decompose the corresponding time-dependent wavefunction in the basis of the eigenstates at $d=0$
\begin{equation}
\label{Eq:d0expand}
|\Psi(t)\rangle=\sum_{nl} c_{nl}(t) |\Psi_{nl}^{(0)}\rangle,
\end{equation}
in analogy to Eq.~\eqref{Eq:Fmulti}. Substituting this into Schr\"odinger equation, we obtain a
set of coupled differential equation for the expansion coefficients $c_{nl}$:
\begin{multline}
\label{Eq:coeffode} i\hbar \dot{c}_{n^\prime l^\prime} =
\sum_{n l} c_{n l}(t) \Big[\Big(E_{n^\prime l'}^{(0)} + \frac 1 2 \mu \omega^2d(t)^2 \Big) \delta_{n,n^\prime}\delta_{l,l^\prime} \\
- \mu \omega^2 d(t) D^{n'l'}_{nl}\Big] .
\end{multline}
Here, $D^{n'l'}_{nl}=\langle \Psi_{n'l'}^{(0)}| z |\Psi_{nl}^{(0)}\rangle$ is the dipole matrix
element, which in the context of the multichannel formalism is defined in
Eq.~\eqref{Eq:MatrixelementMulti}.
We determine the radial part of the single channel wavefunctions and the eigenenergies $E_{nl}^{(0)}$ with the Numerov method \cite{JohnsonNumerov}, using the effective short range phase as a boundary condition at minimal distance. 
From the wavefunctions, one can calculate the matrix elements $D^{n'l'}_{nl}$. Inserting these into Eq.~\eqref{Eq:coeffode}, we are able to solve the equation for the coefficients numerically with standard
routines. Thereby we verify in each case that our basis, limited by the maximal values of the $l$ and $n$
quantum numbers, is large enough and that the results do not change when increasing the basis.

By comparing exact numerical dynamics with the results predicted by the Landau-Zener theory, we
have found, for example, an error of about 0.5\% for the probability of a fast, diabatic passage of the avoided crossing at $d=1$ in the spectrum
of Fig.~\ref{Fig:manualGateSpectrum}, with a speed of 1 mm/s. Similar good agreement was observed in the adiabatic limit, and only for intermediate speeds we found discrepancies of the order of 10\%. This might be due to the complexity of the spectrum, which makes it impossible to isolate an avoided crossing between two eigenstates from the influence of the rest of the eigenstates.
Thus, the Landau-Zener theory is not applicable for quantum gate calculations performed for example in Sec.~\ref{Sec:adiabaticGate}. Also, faster processes lead to excitations to higher vibrational states that cannot be described by the Landau-Zener theory.
\section{Quantum gate \label{Sec:quantum_gate}}
\subsection{Qubit states}

\begin{figure}[ht]
\centering \psfrag{mfa}{\small$m_{f_a}$} \psfrag{mfi}{\small$m_{f_i}$} \psfrag{atom}{\small \bf
$i=\frac32$ Atom ($^{87}$Rb)} \psfrag{ion}{\small \bf $i=\frac32$ Ion ($^{135}$Ba$^+$)}
\psfrag{f=1}{\small$f\!=\!1$} \psfrag{f=2}{\small$f\!=\!2$} \psfrag{-1}{\small$-1$}
\psfrag{-2}{\small$-2$} \psfrag{1}{\small $1$} \psfrag{2}{\small$2$} \psfrag{0}{\small$0$}
\psfrag{brac}{$\Big\{$} \psfrag{bracbig}{$\Big\{$} \psfrag{state1}{\small$|1\rangle$}
\psfrag{state0}{\small$|0\rangle$}
\includegraphics[width=8.6cm]{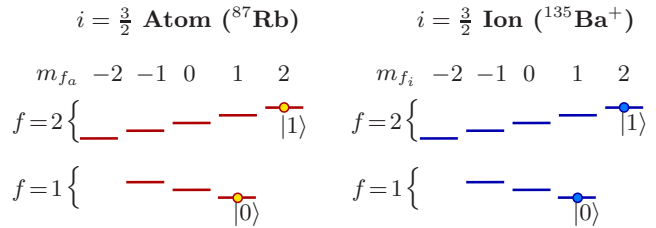} \caption[Qubit states]{(Color online) Specific choice of
the qubit states out of the manifold of hyperfine spin states of a $^{87}$Rb atom and a $^{135}$Ba
$^+$ ion.} \label{Fig:qubits}
\end{figure}

In this Section we apply our model of the spin-state-dependent atom-ion collisions to construct a two-qubit controlled-phase gate. We encode qubit states in hyperfine states of atom and ion. According to our previous
notation, a given two-particle spin state is referred to as a channel $\alpha = \{f_i f_a m_{f_i} m_{f_a} l m_l\}$. The total spin projection $m_f=m_{f_i}+m_{f_a}$ is a conserved quantity during the collision. Therefore two states of unlike
$m_f$ cannot be coupled. In our case it is convenient to pick the computational basis states
\begin{equation}
\begin{split}
|0\rangle_{i,a}&=|f_{i,a}\!=\!1,m_{f_{i,a}}\!=\!1\rangle_{i,a}\\
|1\rangle_{i,a}&=|f_{i,a}\!=\!2,m_{f_{i,a}}\!=\!2\rangle_{i,a}
\end{split}
\end{equation}
according to Fig.~\ref{Fig:qubits}, leading to the two-qubit states
\begin{equation}
\label{Eq:twoQubits}
\begin{split}
|00\rangle&=|f_i\!=\!1,m_{f_i}\!=\!1,f_a\!=\!1,m_{f_a}\!=\!1\rangle=|1,1,1,1\rangle \\
|01\rangle&=|f_i\!=\!1,m_{f_i}\!=\!1,f_a\!=\!2,m_{f_a}\!=\!2\rangle=|1,1,2,2\rangle \\
|10\rangle&=|f_i\!=\!2,m_{f_i}\!=\!2,f_a\!=\!1,m_{f_a}\!=\!1\rangle=|2,2,1,1\rangle \\
|11\rangle&=|f_i\!=\!2,m_{f_i}\!=\!2,f_a\!=\!2,m_{f_a}\!=\!2\rangle=|2,2,2,2\rangle.
\end{split}
\end{equation}
Each of the two-qubit states is represented by a scattering channel. The state $|00\rangle$ has
$m_f=2$ and is coupled to seven other channels, which have higher dissociation energies and therefore they remain closed for $|00\rangle$ collisions. Thus the state $|00\rangle$ is stable with respect to spin-changing collisions.
The channels $|01\rangle$ and $|10\rangle$, belonging
to the $m_f=3$ subspace, are coupled to each other and to two other channels, that are closed for
both $|01\rangle$ and $|10\rangle$ collisions. There is no coupling for the state $|11\rangle$,
since it is the only state in the $m_f=4$ subspace. In this way our choice of the qubit states
minimizes the possibility of spin-changing collisions, and the only process that can lead to
potential losses is the inelastic collision $|10\rangle \rightarrow |01\rangle$.

\subsection{Dynamics of four isolated channels}
\label{Sec:fourChannelDynamics}

Our multichannel theory describes collisions between atom and ion for general spin states of
particles, in particular for the four qubit states introduced in Eq.~\eqref{Eq:twoQubits}. For
simplicity of the numerical calculations, here we do not perform the full multichannel dynamics,
but rather we focus on the regime of applicability of the effective single-channel model, described in
Sec.~\ref{Sec:SingleChannelModel}. For every choice of parameters assumed in our calculations, we
verify that the coupling to other spin states can be neglected. The total Hamiltonian including
external degrees of freedom, for the subspace corresponding to our computational basis, reads
\begin{multline}
\label{Eq:totalHamiltonian} H=H_{00}\otimes |{00}\rangle\langle {00}|+H_{01}\otimes |01\rangle\langle 01|\\
+H_{10}\otimes |10\rangle\langle 10| +H_{11}\otimes |11\rangle\langle 11|.
\end{multline}
We denote a qubit channel by $|A\rangle$ with $A\in\{00,01,10,11\}$. Linear combinations of the
computational basis states form a general two-particle state $|\Psi\rangle=\sum_A
a_A|\Psi^{A}\rangle|A\rangle $, where $|\Psi^{A}\rangle$ denotes the quantum state of the atom-ion
relative motion for the channel $|A\rangle$. Obviously, the time evolution of $|\Psi\rangle$
\begin{equation}
\begin{split}
\label{Eq:4channelTimeEvolution} |\Psi(t)\rangle
=\sum a_A e^{-i E_{A} t/\hbar} |\Psi^{A}(t)\rangle|A\rangle
\end{split}
\end{equation}
is determined by the dynamics of the spatial part of the wave function $|\Psi^{A}(t)\rangle$, which
we evaluate from Eq.~\eqref{Eq:coeffode}. On the other hand, the phases due to the differences in
threshold energies $E_{A}$ in each of the channels, do not lead to state dependent dynamics, and
can be eliminated by single qubit operations (see discussion in the next subsection).

\subsection{Phase gate process}

The two-qubit phase gate is represented by the transformation
\begin{equation}
\begin{split}
\label{Eq:phaseGate}
|00\rangle &\stackrel{         \text{Interaction} }{\longrightarrow} e^{i\phi_{00}}|00\rangle \stackrel{          U_{\mathrm{S}}  }{\longrightarrow} \phantom{e^{i\phi}}|00\rangle,\\
|01\rangle &\stackrel{\phantom{\text{Interaction}}}{\longrightarrow} e^{i\phi_{01}}|01\rangle \stackrel{ \phantom{U_{\mathrm{S}}} }{\longrightarrow} \phantom{e^{i\phi}}|01\rangle,\\
|10\rangle &\stackrel{\phantom{\text{Interaction}}}{\longrightarrow} e^{i\phi_{10}}|10\rangle \stackrel{ \phantom{U_{\mathrm{S}}} }{\longrightarrow} \phantom{e^{i\phi}}|10\rangle,\\
|11\rangle &\stackrel{\phantom{\text{Interaction}}}{\longrightarrow} e^{i\phi_{11}}|11\rangle
\stackrel{\phantom{U_{\mathrm{S}}} }{\longrightarrow}e^{i\phi} |11\rangle,
\end{split}
\end{equation}
performed on the computational basis states. The first step is the controlled interaction of atom an ion that
leads to a specific phase for each two-qubit state. By applying the single-qubit transformation
$U_{\rm S}$ we can undo three of these phases and assign the total gate phase
\begin{equation}
\label{Eq:totalPhase} \phi=\phi_{00}+\phi_{11}-\phi_{01}-\phi_{10}
\end{equation}
to the $|11\rangle$ state \cite{CalarcoIonTraps}. If this phase equals $\pi$, the phase gate, combined with single qubit
gates, is a universal gate for quantum computation, as it is equivalent to a CNOT gate. It is
possible to realize this phase gate scheme within our single-channel model, since the
transformation of each two-qubit basis state can be treated separately.

For our gate scheme, atom and ion are initially prepared in the motional ground state of their respective traps. The channel phases are gained by the control of the relative motion of atom and
ion during the collision. Ideally we aim at obtaining back the motional ground state at the end of the
gate process, so that the phase accumulated by relative motion is assigned to the qubit state.

\subsection{Gate fidelity}

Our definition of the fidelity is based on the overlap of the initial state of relative motion
$|\Psi^A_{\rm ini}\rangle$ with the final state $|\Psi^A(T)\rangle$. In an ideal process these states are equal up
to a state dependent phase. The fidelity needs to account for this phase. For one channel $A$ and
at zero temperature, according to \cite{CalarcoNeutralQuantum} we can define the fidelity $F_A$ as
follows
\begin{equation}
\label{Eq:fidelitydefChannel} F_A=\frac 1 2 \left[1-|\langle \Psi^A_{\rm
ini}|\Psi^A(T)\rangle|\cos(\pi-\Delta \phi_A) \right],
\end{equation}
where  $\Delta \phi_A=\phi_{A}(T)-\phi_A'$ is the difference between the desired channel phase
$\phi_A'$ and the phase $\phi_A(T)$ obtained by actual time evolution. In the following, we will
assume that according to Eq.~\eqref{Eq:phaseGate} the phases for the channels $|00\rangle$,
$|01\rangle$ and $|10\rangle$ are undone perfectly due to the single qubit rotations leading to
$\Delta \phi_{00}=\Delta \phi_{10}=\Delta \phi_{01}\equiv0$, while $\Delta \phi_{11}$ is nonzero.
Hence, for channels $|00\rangle$, $|01\rangle$ and $|10\rangle$ the fidelity is restricted only by
the overlap between initial and final states, while for the state $|11\rangle$ we additionally
require that the total gate phase, computed from the single-channel phases with
Equation~(\ref{Eq:totalPhase}), is $\phi=\pi$.

We can further define the overall gate fidelity as
\begin{equation}
\label{Eq:gateFidelityDef} F_{\mathrm{gate}}=\underset{ A}{\mathrm{min}} \; F_A,
\end{equation}
since in our model the channels are decoupled (we neglect spin changing collisions).

\subsection{Adiabatic regime}

\begin{figure}[ht]
\center \subfloat[]{\label{Fig:manualGateSpectrum}\includegraphics[width=3.8cm]{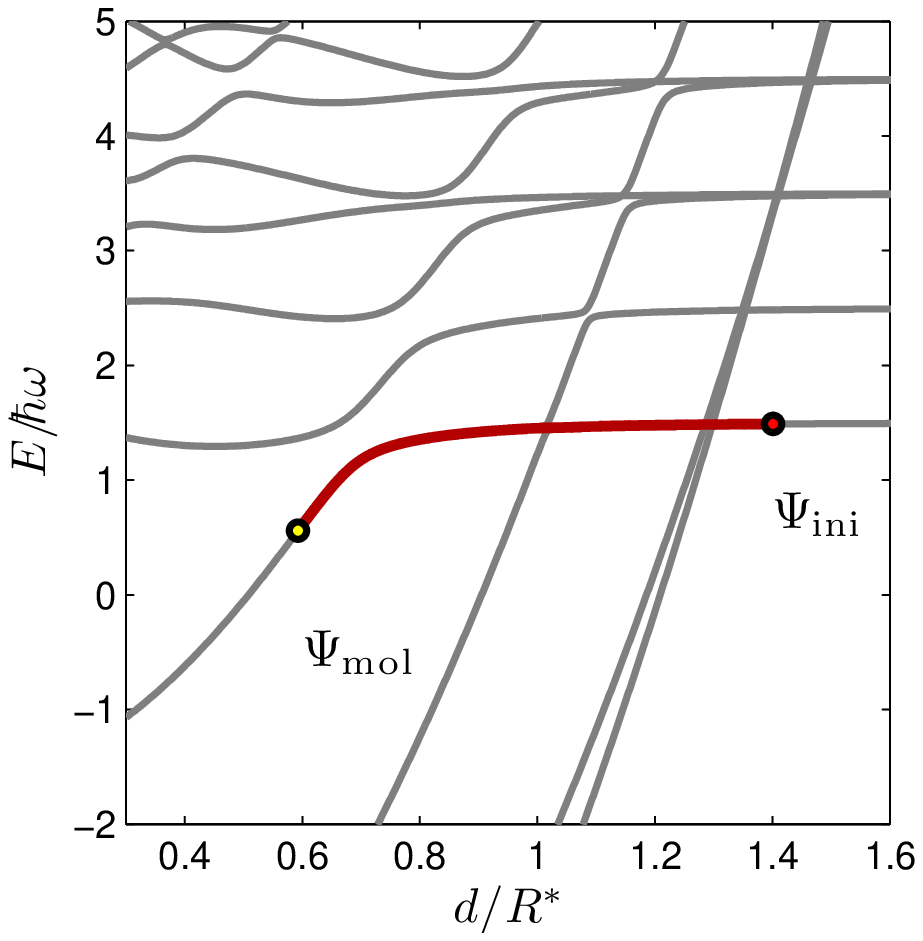}}\hspace{0.1\linewidth}
\subfloat[]{\label{Fig:manualGateZoom}\includegraphics[width=3.8cm]{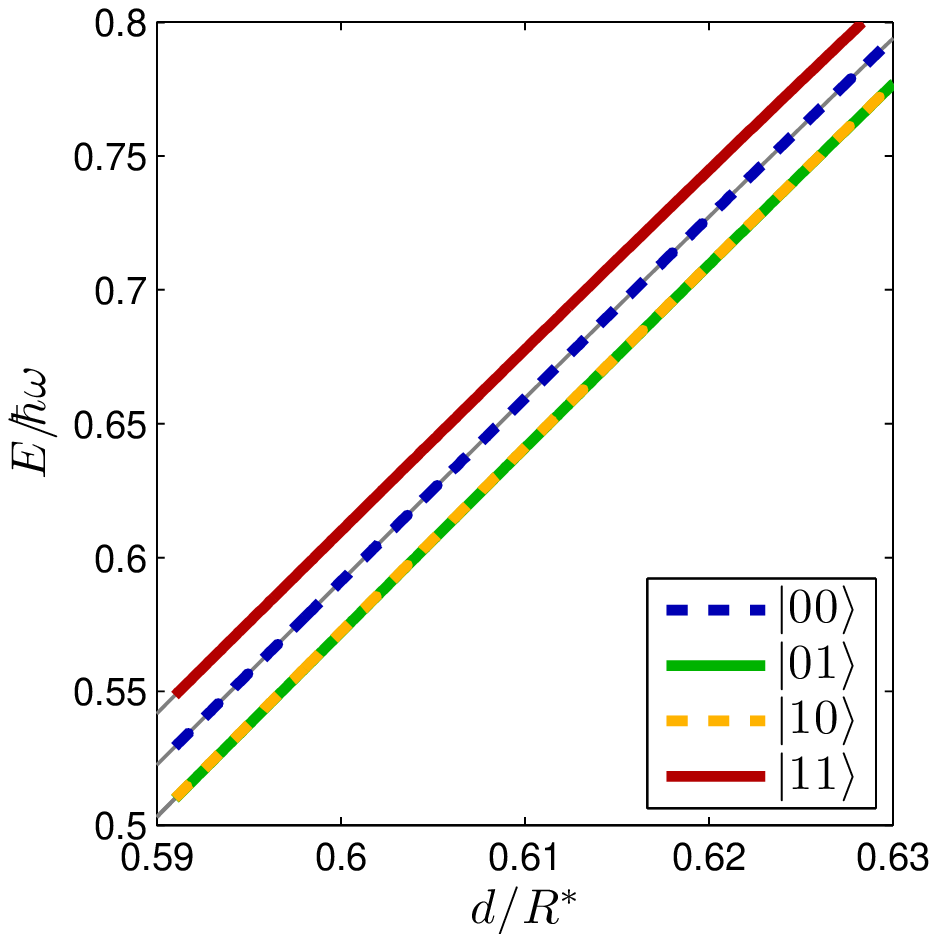}}
\caption[Energy Curve in the Adiabatic Gate Process]{(Color online) Correlation diagrams for
$a_s=0.90$ and $a_t=0.95$, where for each of the qubit pairs we subtract the threshold energy of the corresponding channel. The energy curve followed in the adiabatic process is
marked with a thick, red (dark gray) line in the left Figure (a). We only show the complete diagram for the $|11\rangle$-channel. Small energy differences of
the channels around $d=d_{\rm min}$ can be seen in the close-up (b). These differences
are the basis of our proposal for realizing an atom-ion phase gate.} \label{Fig:manualGate}
\end{figure}

The adiabatic dynamics can be understood with the help of the correlation diagrams introduced in Sec.~\ref{Sec:CorrDiag}. Our gate scheme aims at an adiabatic transfer from an initial oscillator state $\Psi_{\rm{ini}}$ to a molecular state $\Psi_{\rm mol}$, and back to the initial state.
This is achieved by a variation of the trap distance across an appropriate avoided crossing, which we choose after investigating the correlation diagram.
For example, the resonance at the trap distance $d\sim0.7$ in Fig.~\ref{Fig:manualGateSpectrum} appears strong enough and we use it in numerical calculations in the following.
During the transfer process each logical basis state acquires a different phase, since the energies of molecular states depend on the channel (see Fig.~\ref{Fig:manualGateZoom}).
The phase accumulated for each channel in an adiabatic transfer process is given by the integral
\begin{equation}
\label{Eq:potentialPhase} \phi^{A}_{\mathrm{pot}}=-\frac 1 {\hbar}
\int_{t_\mathrm{min}}^{t_{\mathrm{max}}} E^A(t) \ud\,t,
\end{equation}
where $E^A$ is the energy of the adiabatic eigenstate depicted as a function of $d$ in Fig.~\ref{Fig:manualGate} (thick curve). In the adiabatic regime we avoid excitations to higher vibrational states by keeping the velocity $\dot{d}$ small,
compared with the characteristic velocity of the harmonic motion in the trap: $\dot{d} \ll
\sqrt{\hbar \omega/\mu}$. On the other hand we choose the velocity across weaker
resonances at larger distances high enough to pass them diabatically, as seen in Fig.~\ref{Fig:manualGate}.  We want to find a particular function $d(t)$, which, if applied on the trapped atom-ion system, results in a desired total gate phase, while ensuring diabatic passage of the weak resonances as well as adiabatic passage of the strong resonance. Since the total phase depends on the difference between single-channel phases (see
Eq.~\eqref{Eq:totalPhase}), the gate speed, in fact, is determined by the differences in the
potential energy curves of unlike channels.

For the sake of concreteness we assume specific values of the singlet and triplet scattering
lengths, in such a way that our single-channel effective model is applicable. For our calculations we choose $a_s=0.90R^*$ and $a_t=0.95R^*$. According to the procedure described in Sec.~\ref{Sec:SingleChannelModel} the estimate of the error introduced by the model is $2\times10^{-3}$. For singlet and triplet scattering
lengths that differ by more than $10\%$ the channel mixing becomes already significant and does not allow for a single-channel description.

Actually, the singlet and triplet scattering lengths are uniquely determined by the specific choice
of the atom-ion system we describe. So far these parameters have not been measured experimentally, for any atom-ion system.
However, as soon as the accurate values of $a_s$ and $a_t$ are determined, one can repeat the calculations
with the physically correct parameters, which may require going beyond the single-channel model, and including the full multichannel dynamics according to
Sec.~\ref{Sec:multichannelQDT}.

Assuming the single-channel effective model we first compute a correlation diagram for each of the
channels. This is done by diagonalizing the Hamiltonian in the basis of eigenstates evaluated at
$d=0$. The result is depicted in Fig.~\ref{Fig:manualGateSpectrum} for the $|11\rangle$ channel.
The diagrams show small differences in the molecular states at small distances
(Fig.~\ref{Fig:manualGateZoom}), since the energy of molecular states depends on the atom-ion spin
configuration (the qubits).

\subsubsection{Numerical simulation of an adiabatic phase gate }
\label{Sec:adiabaticGate} In our simulation of the gate process the initial and final trap
separations coincide: $d(0)=d(T)=d_{\rm max}$. We assume that initially the atom and ion are each in the
ground-state of its own trap. The distance $d_{\rm max}$ is determined in such a
way that
there are no bound states in the vicinity of $d_{\rm max}$ that would influence the
harmonic-oscillator ground-state.

The controlled time evolution of our atom-ion system requires the appropriate adjustment of the
distance $d(t)$ between the two trapping potentials as a function of time. The slope of this
function is essential for the result.

\begin{figure}[ht]
\subfloat[]{\label{Fig:adiabaticdminVariation_d_t}\includegraphics[width=4cm]{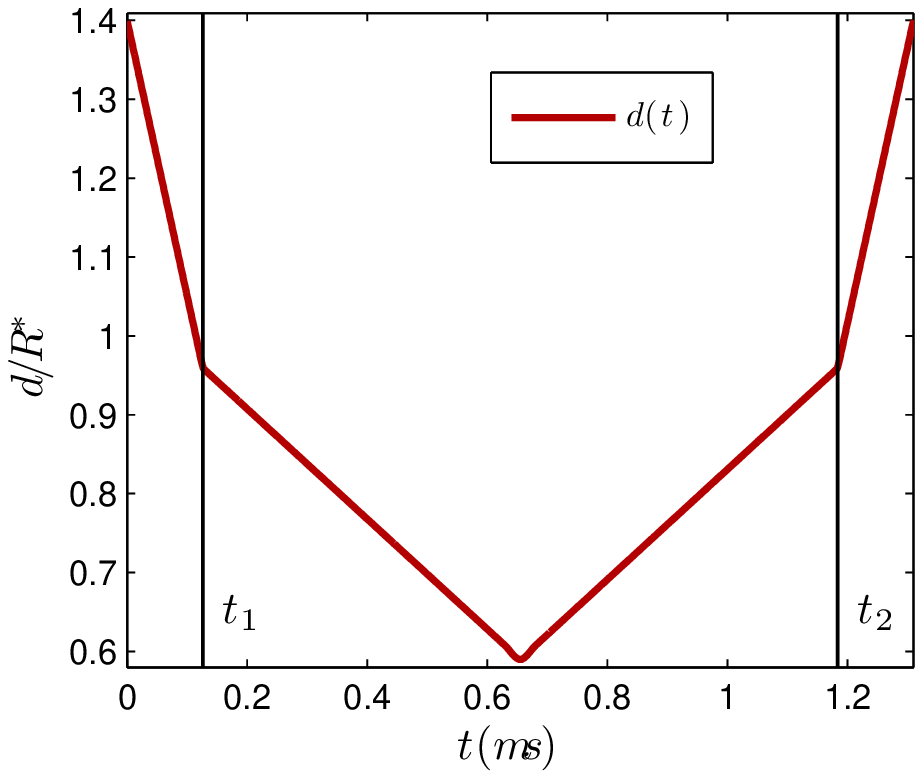}}
\subfloat[]{\label{Fig:adiabaticdminVariation_ph}\includegraphics[width=4cm]{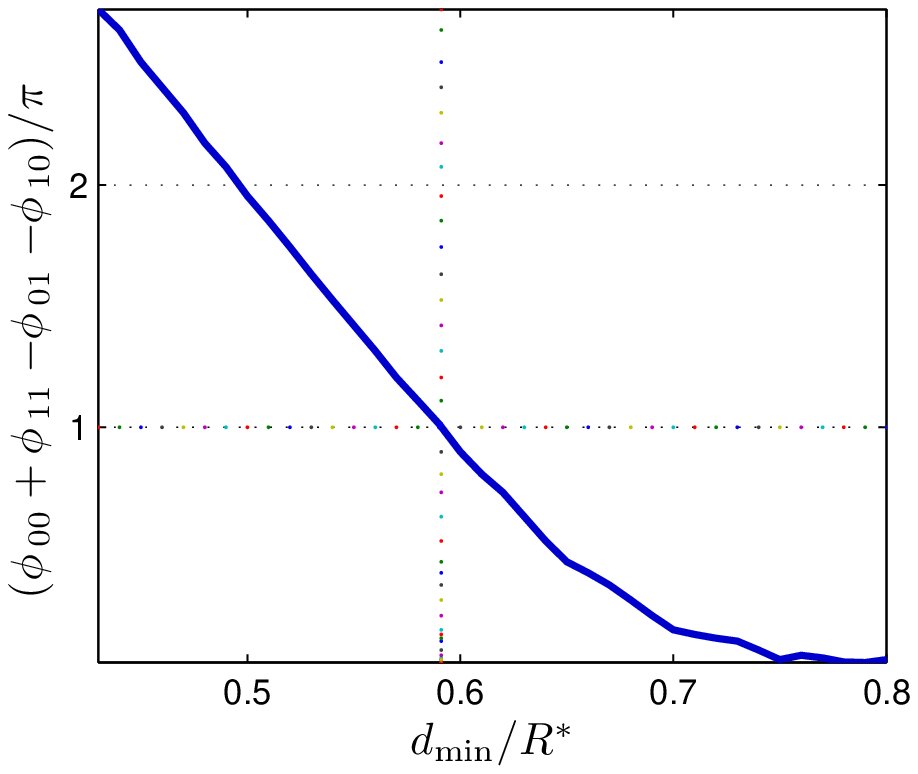}}
\caption[Adiabatic Gate Function]{(Color online) (a): The function $d(t)$ starts at $d_{\rm
max}=1.4R^*$ with a slope of $d'_1=0.5R^*/(\hbar/E^*)$ large enough to diabatically pass weaker resonances. The velocity is changed to around $d=0.95R^*$ to $d'_2=0.1R^*/(\hbar/E^*)$ in order to ensure an adiabatic traversal of a stronger resonance. The system is brought to the initial distance with the reversed pulse and the kinks are smoothed to avoid motional excitations.
The characteristic unit of speed is $R^*/(\hbar/E^*)= 2.05$mm$/$s.
(b): gate phase as a function of $d_{\rm min}$ using the described $d(t)$ shape.
We find that with $d_{\rm min}=0.591$ a gate phase of $\phi=1.009\pi$ is reached.
} \label{Fig:adiabaticdminVariation}
\end{figure}

%
%
In order to follow the energy curve depicted in
Fig.~\ref{Fig:manualGateSpectrum}, we construct a specific function $d(t)$. 
We start at $d_{\rm max}$ with an initial velocity $d'_1=0.5R^*/(\hbar/E^*)$,
which is large enough to traverse weaker resonances diabatically.
On the other hand, much larger velocities would cause unwanted motional excitations in the trap.
At $d=0.95R^*$, the velocity is decreased to $d'_2=0.1R^*/(\hbar/E^*)$ in order to adiabatically convert the trap state into a molecular bound state using a stronger resonance.
The curve is followed down to some minimal distance $d_{\rm min}$. Then, the reversed pulse brings the system to the initial trap separation. Fig.~\ref{Fig:adiabaticdminVariation_d_t} shows the complete $d(t)$ function.
It is known that sharp kinks can cause motional excitations,
therefore we use a smooth function 
$d(t)=\tilde{d}+1/2(d'_1+d'_2)t\pm\sqrt{(d'_1-d'_2)^2t^2+C^2}$ to change between two slopes $d'_{1,2}=0.5$ and $0.1 R^*/(\hbar/E^*)$. Here, the $+(-)$ sign is used for increasing(decreasing) slope, $\tilde{d}$ is an offset and $C$ is a parameter adjusting the curvature at the kink. Around the turning point at $d_{\rm min}$ $d(t)\sim t^2$.

\begin{figure}[ht]
\center \includegraphics[width=8.6cm]{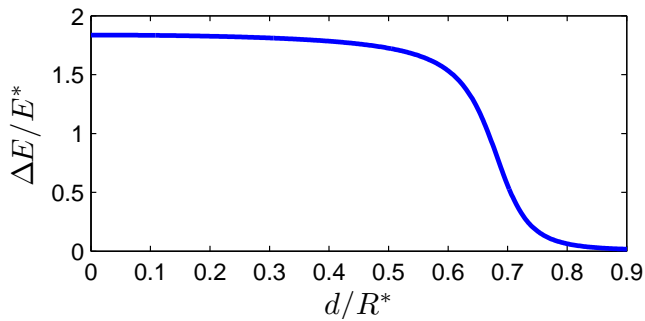}
\caption[Energy Splitting]{(Color online) Energy difference as a function of trap separation.
For $d\lesssim 0.3R^*$ the function is practically constant, which,
for our phase gate process, means that bringing the traps closer than $d_{\rm min}\sim 0.3R^*$ does not lead to a significant advantage.}
\label{Fig:dmin_dE_variation}
\end{figure}

\begin{figure}[ht]
\subfloat{\includegraphics[width=8.6cm]{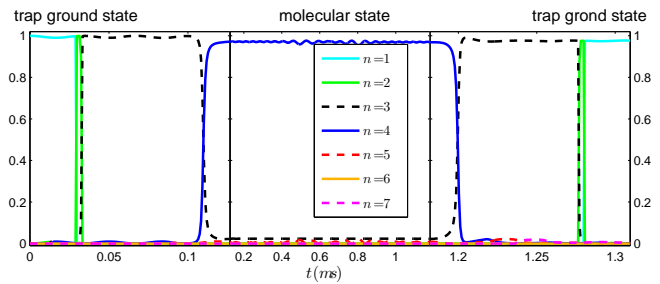}}\\
\caption[Populations of Energy Eigenstates for the Adiabatic Gate]{(Color online) Population of the most important
adiabatic eigenstates for the qubit channel $|11\rangle$ during the adiabatic gate process (the remaining channels show similar behavior). States are labeled according to their energy in the correlation diagram in ascending order. The initial state at $t=0$ is the relative motion ground state in the trapping potential for $d=d_{\mrm{max}}$. The label of this state is set to $n=1$. The molecular state marked in Fig.~\ref{Fig:manualGateSpectrum} then bears the label $n=4$. Around a crossing, the labels of molecular/trap states switch because their energies change order. Our goal is to follow the thick red (dark gray) curve in Fig.~\ref{Fig:manualGateSpectrum}. We observe that indeed two resonances are passed diabatically. At $t=T/2$ the state $n=4$ is reached with relatively high fidelity, which means that a molecular ion is formed here. At $d(T)=d_{\rm max}$ we regain the
initial state again (the state dependent phase of the final state is not depicted
here). To better show the features of the curves, the time axis is squeezed between $t_1=0.12$ ms and $t_2=1.18$ ms, where the velocity is lower (see Fig.~\ref{Fig:adiabaticdminVariation}).}
\label{Fig:adiabaticGatePopulation}
\end{figure}

We can now compute the solution of the Schr\"odinger
equation numerically at a given time by solving Eq.~\eqref{Eq:coeffode} with standard numerical
routines \comment{(we use the Matlab routine \textit{ode45})}.
This yields the gate phase which, for example, can be adjusted by a variation of $d_{\rm min}$. This phase is in fact a phase
difference accumulated due to the energy splitting
\begin{equation}
\label{Eq:deltaE}
 \Delta E=E_{\rm mol}^{00}+E_{\rm mol}^{11}-E_{\rm mol}^{01}-E_{\rm mol}^{10}
\end{equation}
shown in Fig.~\ref{Fig:dmin_dE_variation} as a function of trap separation; the larger $\Delta E$, the faster a phase difference is reached. Thus, decreasing $d_{\rm min}$ increases the phase, as seen in Fig.~\ref{Fig:adiabaticdminVariation_ph}. We find a gate phase of $\phi=1.009\pi$ at $d_{\rm min}=0.591R^*$.
The corresponding gate fidelity is $F_{\rm gate}=0.994$ according to Eq.~\eqref{Eq:gateFidelityDef}. The process takes the time $T=9.14\hbar/E^*$, which equals $T=1.31$ms for our choice of the trapping frequency $\omega=2\pi\times30$kHz and the masses and hyperfine structure of the $^{87}$Rb atom and the $^{135}$Ba$^+$ ion. We show the population of the instantaneous eigenstates in Fig.~\ref{Fig:adiabaticGatePopulation}.
At $d_{\rm max}$ the trap ground state is labeled with $n=1$. The system is initialized in this state. At half gate time the quantum state changed to $n=4$, which is the molecular state at $d_{\rm min}$. Finally, at the end of the gate process, we almost perfectly obtain back the initial state.


\subsection{Fast gate using optimal control}

Quantum optimal control techniques are a powerful tool that allows to increase the fidelity of a
time evolution process by finding an appropriate pulse shape for some external control parameter. The
outcome of the controlled collision of an atom and an ion is very sensitive to the particular shape
of the time dependence in the trap distance $d(t)$. It is hard to manually design a specific function
$d(t)$ that leads to a phase gate fidelity very close to unity.

The most significant problem in our specific example calculation results from a relatively weak
resonance at $d_0\approx1R^*$, which we want to pass in a non-adiabatic way. We could not find an
optimal constant slope that brings us from the trap state at $d>d_0$ to the trap state with $d<d_0$ across the resonance without losses. Large velocities lead to excitations of energetically higher states, while the
consequence of smaller velocities is a non-negligible population of the molecular state that crosses the trap
state. This population is not fully recovered on the way back. In Sec.~\ref{Sec:adiabaticGate} we
nevertheless found a process that yields a gate fidelity of $F_{\rm gate}=0.994$.

With optimal control we can not only find a pulse shape that produces a satisfactory gate fidelity, but also go beyond the adiabatic regime and reduce the gate time. By applying larger velocities, we allow for excitations to
higher energy levels. Making use of interference effects, an appropriate $d(t)$ pulse shape can
undo these excitations in the final state of the process.

This optimal pulse shape is found here with an iterative optimization algorithm called intermediate
feedback control, which is for example introduced in \cite{CalarcoFeshbach}. We start with an
initial guess for the control function $d(t)$, which in general does not yield a satisfactory
fidelity. The fidelity of the process is increased in every iteration step by updating $d(t)$. We
divide the time axis in small time steps $dt$. At each time step we evolve the wave function
forward in time using the Crank-Nicholson scheme \cite{NumericalRecipes}. The update of $d(t)$ is
successively performed in every time step.

\subsubsection{Enhancing the fidelity of the adiabatic gate}
The adiabatic gate process of Sec.~\ref{Sec:adiabaticGate}  has the fidelity $F_{\rm gate}=0.994$.
With only three iterations of the optimal control algorithm we can enhance this fidelity to $F_{\rm
gate}=1-7\times10^{-4}$ \comment{or 0.99929}, which even exceeds the validity of the underlying
single-channel model. The gate phase is improved to $1.0026\pi$. The optimized function $d(t)$
shows small scale variations ('wiggles') that are a typical feature of the used optimization
algorithm. These wiggles have the amplitude $\sim0.002 R^*\approx 0.6$nm and happen on a time scale
of the order of $ 10$ \hbox{\textmu}s. This amplitude is smaller than the uncertainty of the ion trap center position in up-to-date experimental realizations.

\subsubsection{Fast Gate Scheme}
It is desirable to reduce the gate time to a minimum. In our case this minimum is given by the time
that is at least required for accumulating the gate phase $\phi=\pi$. We profit from the energy
differences of the molecular states for the different channels. The differences are largest at
small distance $d_{\rm min}$, but below $d\approx 0.3R^*$ they are practically constant.
The fastest possible gate needs to effectively transport the atom-ion relative
wavefunction into a molecular state, where the gate phase is accumulated during a certain time. The gate ends with a reversed pulse and brings the system to the initial trap ground state, while the logical phase is preserved.

\begin{figure}[ht]
\subfloat[\label{Fig:d_t_opt}]{\includegraphics[width=8.6cm]{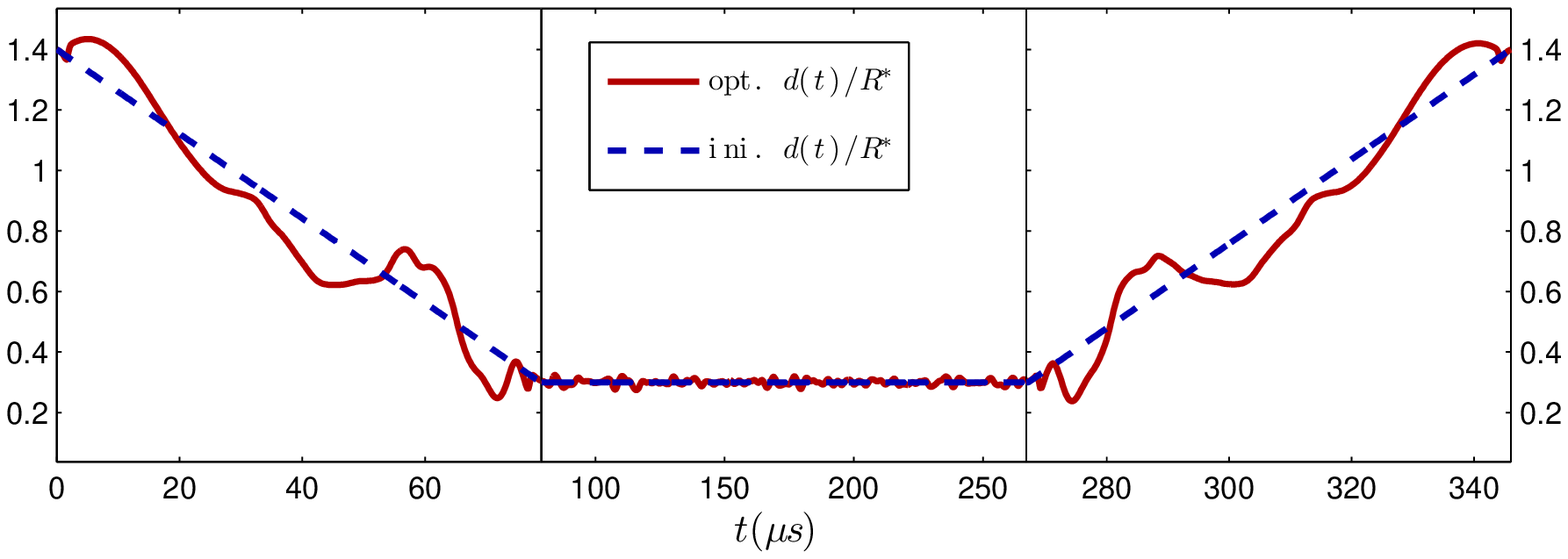}}\\
\subfloat[\label{Fig:fastGateb}]{\includegraphics[width=8.6cm]{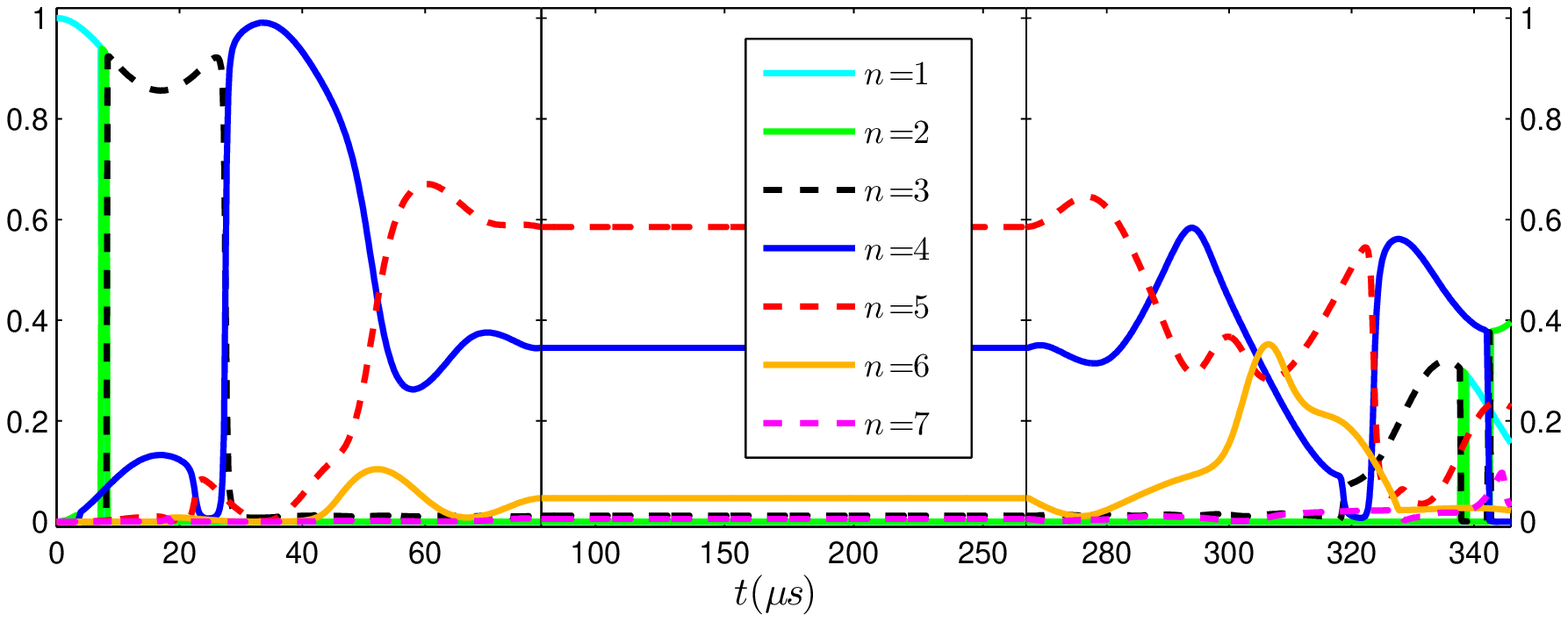}}\\
\subfloat[\label{Fig:fastGatec}]{\includegraphics[width=8.6cm]{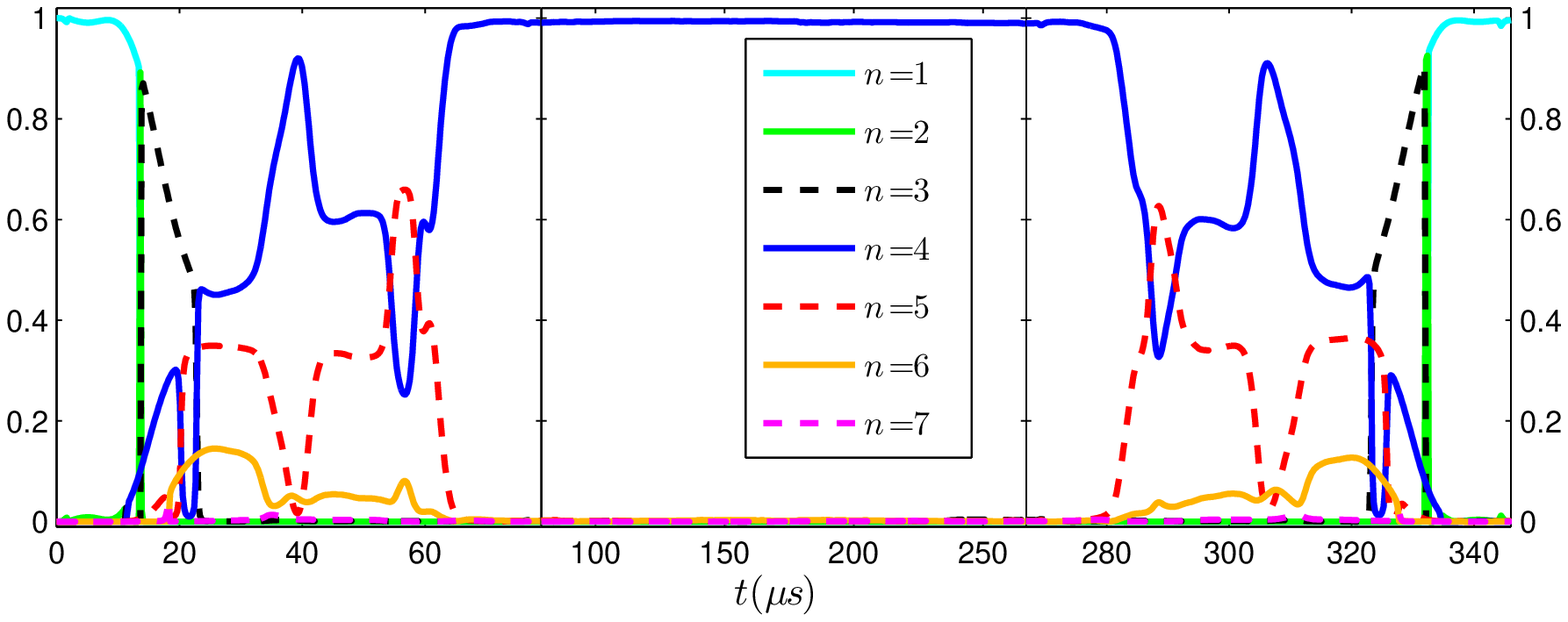}}
\caption[Fast Gate Process]{(Color online) (a): Optimized $d(t)$ function and initial guess (dashed) for the
fast gate process. The initial state is the trap ground state at $d_{\rm max}$. The populations of
the adiabatic eigenstates are changed by the optimization process. Applying the initial guess pulse
higher vibrational and also molecular states become populated (b). These excitations are
prevented or undone using the optimized pulse (c). In this case, at $d_{\rm min}$ the desired molecular
state $n=4$
is reached almost perfectly. For simplicity we only show the plots for the qubit
channel $|11\rangle$ here, the situation is very similar for the other channels. Note the different
time scales used for the transport and phase-accumulation sequences respectively.} 
\label{Fig:fastGatePopulationsAll}
\end{figure}

The optimal control algorithm can be used to build such a gate process. As a first step we design
$d(t)$ changing the quantum state of atom and ion from the trap ground state at $d_{\rm
max}=1.4R^*$ to the molecular state at $d_{\rm min}=0.3R^*$, for all channels during the transport time
$t_{\rm trans}$. The optimization objective $J=\sum_A 2 \mrm{Re}\left\{ \langle\Psi^{A}(t_{\rm
trans})|\Psi^A_{\rm mol}\rangle\right\}$ for this step aims at maximizing the overlap of the
time-evolved state with the desired molecular state. The second step is a stationary evolution at $d_{\rm min}$,
where the main part of the differential phase is accumulated. Subsequently we perform the reverse of the initial pulse. The combined $d(t)$ pulse is shown in Fig.~\ref{Fig:d_t_opt}.

The total gate time is $T=346$\hbox{\textmu}s.
With respect to the adiabatic case this means a reduction by a factor of 4.
Since our scheme makes use of
the channel energy differences in the molecular state at $d=d_{\rm min}=0.3R^*$, we can estimate a
quantum speed limit for this process $ T_{\rm limit}= \pi \hbar / \Delta E$, with $\Delta E$ from Eq.~\eqref{Eq:deltaE}, which gives the minimal gate time, neglecting transport durations and infidelities of the molecular
state's population. In our example this limit is $T_{\rm limit}\approx 250$\hbox{\textmu}s. Our gate
process time lies very close to this value, considering an overall transport time of $2*t_{\rm
trans}=158$\hbox{\textmu}s. We note that a part of the phase is
accumulated during the transport phase, since we already enter the regime where the energy differences become
significant.

Further reducing the transport time may be possible, but the optimization algorithm we used stopped
converging in reasonable time for larger transport velocities. However, we already could
significantly accelerate the process utilizing optimal control techniques.

\subsubsection{Perspectives for further improvement}

Certainly the
gate speed would improve if the difference between singlet and triplet
scattering lengths was larger than assumed in our example calculation. In general case one can use our multichannel formalism for an accurate description of
the dynamics, beyond the single-channel approximation. We point out that the essential parameters $a_s$ and $a_t$ are still unknown and they need to be measured in experiments 
in order to do realistic calculations for specific systems.
However, our work demonstrates the feasibility of an ion-atom quantum gate even based on the simplified scheme we assume for calculations.

Further possibilities occur for very different values
$a_s$ and $a_t$. In this case especially the $|10\rangle$ and $|01\rangle$ states are
coupled strongly. In this case optimal control mechanisms
can be used to suppress effects of spin-changing collisions in the final
state, in order to realize a phase
gate. The coupling of $|01\rangle$ and $|10\rangle$ could also be effectively
used for a SWAP gate, which inverts the populations of
these two channels---or for its square root, which in combination with single-qubit rotations constitutes an alternative universal set of gates for quantum computation.
\section{Conclusions and outlook}
\label{Sec:Outlook}

In this work we analyzed the spin-state-dependent interaction between a single atom and a single
ion guided by external trapping potentials. We applied our insight on this system in order to
realize a two-qubit quantum gate process and thereby provide the basic ingredients for quantum
computation with atoms and ions combined in one setup. This work was motivated by recent
experimental possibilities combining magneto-optical traps or optical lattices for atoms, and RF
traps for ions. These experiments are currently established in several groups
worldwide \cite{BaRbI,Vuletic}.

We started our description of controlled interaction of an atom and an ion by formulating a
multichannel quantum-defect theory for trapped particles, analogous to the free-space case
discussed in \cite{ZbyszekMQDT}. This step simplifies the description of atom-ion collisions as it does not require a detailed knowledge of the molecular potentials at short range. Experiments measuring the positions of Feshbach
resonances can determine the essential two parameters for our theory---the singlet and triplet
scattering lengths. Since these experiments have not yet been performed, here, we have focused on the case of similar singlet and triplet scattering lengths given in order of magnitude by $R^\ast$, while the general case of different scattering lengths have been discussed only  qualitatively.

We were able to reduce the multichannel formalism to an effective single-channel model that singles
out a specific spin state of atom and ion. This model is found to be accurate for similar values of
the singlet and triplet scattering lengths. In calculations we have assumed $a_s = 0.9 R^{\ast} = 4989 a_0 $ and $a_t = 0.95 R^{\ast} =5266 a_0$ for the singlet and triplet scattering lengths, respectively. In contrast, opposite signs of scattering lengths exclude a single-channel description. We estimated
the error introduced by our specific single-channel model to $2\times10^{-3}$, which is due to a
mixing of channels in the eigenstates of the system. Taking even closer values of singlet and triplet scattering lengths lead to a better applicability of the single-channel description.

Where applicable, our effective single-channel model can be implemented in calculations in the context of ultracold chemistry as well as ultracold scattering physics. A single-channel description assigning
quantum-defect parameters to each channel separately was already discussed in \cite{ZIAtomIon}. However, the spin state was not included in previous research.
In the present approach, starting from the fundamental parameters $a_s$ and $a_t$ of the multichannel formalism, we derive the quantum-defect parameters of each isolated channel consistently. We take the channel coupling into account and we estimate the error introduced by assuming isolated channels.
Therefore, we can apply the model to quantum computation schemes
that store qubits in internal spin states of atom and ion.

A remarkable feature of the system is trap-induced shape resonances that couple molecular bound
states to unbound trap states. Quasistatic eigenenergy curves show these resonances as avoided
crossings. They can be used to form ultracold trapped molecular complexes and thereby allow full
control of cold chemical reactions.

Trap-induced resonances form a basis for our idea of the phase gate process as well. Initially an
atom and an ion are prepared in the trap vibrational ground state. We realize a qubit-dependent
two-particle phase via controlling the external degrees of freedom. By bringing the traps close
together we let the particles interact and finally separate them, obtaining the motional ground
state again. In doing so we cross weaker resonances diabatically (remaining in a trap state) and
then follow a stronger resonance adiabatically into a molecular bound state, where a two-qubit phase is
accumulated. Since the positions of the resonances are different for each spin combination, the
accumulated phase is different for each qubit channel and we are able to control the trap distance
in such a way that a two-qubit phase gate is realized. This phase gate, in combination with single
qubit rotations, is a universal gate for quantum computation.

We performed numerical simulations of the controlled collision specifically for a $^{135}$Ba$^+$
ion interacting with a $^{87}$Rb atom, each guided by a spherically symmetric harmonic trap with
$\omega_{i,a}=2\pi\times30$kHz.  We have chosen specific hyperfine-qubit states to obtain the four
qubit-channels $00,01,10,11$. In this framework we developed a two-qubit phase gate process entangling
atom and ion. We thereby showed that trap-induced resonances can be used to control the interaction
of atom and ion. The error for our gate process is $1\times10^{-3}$ and in this case the gate
time is $1.3$ ms. Using optimal control techniques we were able to accelerate the process to
$346$\hbox{\textmu}s. In future work we plan to decrease the gate time by using higher trapping
frequencies allowing faster transport.

Our choice of very similar scattering lengths allowed to perform single-channel calculations, but
on the other hand it limits the gate time, as the energy splitting between the channels is rather
small. A more general description can be done in the framework of our multichannel
formalism, allowing for arbitrary combinations of singlet and triplet scattering
lengths. It is possible that the actual values of the scattering lengths are in fact very different,
which would require a more complicated multichannel computation, but also possibly allows much
faster quantum gates. However, already in the regime we considered, a gate time below a millisecond is demonstrated.

%
In our model, we assumed harmonic trapping potentials, and the oscillator frequencies were equal
for atom and ion. Among the advantages of ions for quantum computation is the existence of much
tighter trapping potentials than for atoms. The basic ideas developed in this paper are expected to
be applicable to such more general situations with different trapping frequencies. A generalization
is highly desirable, but will lead to a more complicated theoretical treatment, as for example
center-of-mass motion becomes coupled to the relative motion. Very elongated cigar-shaped traps
have already been treated in previous works.
One of our goals is the consideration of particular experimental
realizations in order to describe them with our theory, compare the results or suggest directions
of experimental research.

In the present work we did not use external magnetic fields in order to manipulate the interaction.
Magnetic induced Feshbach resonances have been applied very successfully to engineer neutral atom
collisions. Future investigations will include magnetic fields to control the atom-ion interaction
even more efficiently and possibly combine trap-induced resonances and Feshbach resonances for this
purpose. Our work can be seen as a principle investigation of a new, interesting physical system
and can be extended in many directions.

\begin{acknowledgments}
We acknowledge support by the EU under the Integrated Project SCALA,  the German Science Foundation through SFB TRR21 Co.Co.Mat,
the National Science Foundation through a grant for the Institute for Theoretical Atomic, Molecular and Optical Physics at Harvard
University and Smithsonian Astrophysical Observatory,
and the Polish Government Research Grant for the years 2007-2010.
The authors would like to thank P. Zoller, P. Julienne, J. Hecker Denschlag and P. Schmidt for fruitful discussions. HDB thanks M. Trippenbach and the University of Warsaw for very rewarding two-week visit.
\end{acknowledgments}

\end{document}